# Linear Mixed Models with Marginally Symmetric Nonparametric Random Effects

Hien D. Nguyen and Geoffrey J. McLachlan

February 14, 2016

School of Mathematics and Physics, University of Queensland

## Abstract

Linear mixed models (LMMs) are used as an important tool in the data analysis of repeated measures and longitudinal studies. The most common form of LMMs utilize a normal distribution to model the random effects. Such assumptions can often lead to misspecification errors when the random effects are not normal. One approach to remedy the misspecification errors is to utilize a point-mass distribution to model the random effects; this is known as the nonparametric maximum likelihood-fitted (NPML) model. The NPML model is flexible but requires a large number of parameters to characterize the random-effects distribution. It is often natural to assume that the random-effects distribution be at least marginally symmetric. The marginally symmetric NPML (MSNPML) random-effects model is introduced, which assumes a marginally symmetric point-mass distribution for the random effects. Under the symmetry assumption, the MSNPML model utilizes half the number of parameters to characterize the same number of point masses as the NPML model;

thus the model confers an advantage in economy and parsimony. An EM-type algorithm is presented for the maximum likelihood (ML) estimation of LMMs with MSNPML random effects; the algorithm is shown to monotonically increase the log-likelihood and is proven to be convergent to a stationary point of the log-likelihood function in the case of convergence. Furthermore, it is shown that the ML estimator is consistent and asymptotically normal under certain conditions, and the estimation of quantities such as the random-effects covariance matrix and individual *a posteriori* expectations is demonstrated. A simulation study is used to illustrate the gains in efficiency of the MSNPML model over the NPML model under the assumption of symmetry. A pair of real data applications are then used to demonstrate the manner in which the MSNPML model can be used to draw useful statistical inference.

# 1 Introduction

Linear mixed models (LMMs) are used as a fundamental tool for the statistical analysis of longitudinal data and data with repeated measurements; see McCulloch and Searle (2001, Ch. 6), Pinheiro and Bates (2000, Ch. 1), and Verbeke and Molenberghs (2000) for introductions on the topic. In the style of Laird and Ware (1982), an LMM can be characterized as follows.

Let $\boldsymbol{Y}_j = \left(Y_{j1}, ..., Y_{jn_j}\right)^T$ be a vector of $n_j$ responses belonging to individual $j$, for $j = 1, ..., n$, where $n$ is the total number of individuals. Further, let each measurement $Y_{jk}$, for $k = 1, ..., n_j$, be dependent upon covariate vectors $\boldsymbol{x}_{jk} \in \mathbb{R}^p$ and $\boldsymbol{z}_{jk} \in \mathbb{R}^q$, and for each $j$, let $\boldsymbol{B}_j \in \mathbb{R}^q$ be a vector of individual random effects arising from the *a priori* probability distribution $F_{\boldsymbol{B}}(\boldsymbol{b})$ with density $f_{\boldsymbol{B}}(\boldsymbol{b})$. We say that the data arises from an LMM if for each $j$ and $k$,

$$Y_{jk}|\left(\boldsymbol{B}_j = \boldsymbol{b}_j\right) = \boldsymbol{x}_{jk}^T\boldsymbol{\beta} + \boldsymbol{z}_{jk}^T\boldsymbol{b}_j + E_{jk}, \tag{1}$$

where $\boldsymbol{\beta} \in \mathbb{R}^p$ is a vector of fixed effects and $E_{jk}$ is a random error with probability density $f_E(e)$. Here, a superscript $T$ indicates matrix transposition.

The main difficulty that arises in the use of LMMs is the evaluation and manipulation of the marginal density of $\boldsymbol{y}_j$, which has the general form

$$f_{\boldsymbol{Y}_j}(\boldsymbol{y}_j) = \int_{\mathbb{R}^q} \left[ \prod_{k=1}^{n_k} f_E\left(y_{jk} - \boldsymbol{x}_{jk}^T \boldsymbol{\beta} - \boldsymbol{z}_{jk}^T \boldsymbol{b}_j\right) \right] f_{\boldsymbol{B}}(\boldsymbol{b}_j) \, \mathrm{d}\boldsymbol{b}_j, \qquad (2)$$

and can often be quite complex due to the integration form.

The traditional approach in dealing with the complexities of (2) is to set the error density as

$$f_E(e) = \phi\left(e; 0, \sigma^2\right), \qquad (3)$$

where

$$\phi\left(e; \mu, \sigma^2\right) = \frac{1}{\sqrt{2\pi\sigma^2}} \exp\left[-\frac{(e-\mu)^2}{2\sigma^2}\right]$$

is the normal density function with mean $\mu$ and variance $\sigma^2$, and to let $f_{\boldsymbol{B}}(\boldsymbol{b})$ be a multivariate normal density. This approach results in $f_{\boldsymbol{Y}_j}(\boldsymbol{y}_j)$ having the form of a multivariate normal density function, and allows for simple inference by maximum likelihood (ML) estimation via an expectation–maximization (EM) algorithm; see Laird and Ware (1982, Sec. 4) and McLachlan and Krishnan (2008, Sec. 5.9) for details.

It is well known that the estimation of the fixed effects $\boldsymbol{\beta}$ is robust to the specification of the random-effects distribution. However, this robustness does not extend to the characterization of the random effects in the case of misspecification. The robustness as well as the effects of misspecification are explored in Agresti et al. (2004), Butler and Louis (1992), and McCulloch and Neuhaus (2011). For instance, in all three articles, the authors note that the estimates for the fixed effects tended not to be influenced greatly by the choice of the random-effects model. However, Agresti et al. (2004) note that the usual normal model can be highly inefficient when the true random-effects model is polarizing, such as in the case of binary or mixture random-effects distributions. It is further remarked in McCulloch and Neuhaus (2011) that the estimation of random intercept coefficients can be biased by making incorrect assumptions regarding the shape of the underlying random-effects distribution.

It is well known that the specification of the random-effects density has little effect on the estimation of the fixed effects $\boldsymbol{\beta}$. However, the same cannot be said regarding estimation and characterization of the random effects in general. see for instance Agresti et al. (2004), Butler and Louis (1992), and McCulloch and Neuhaus (2011).

Multiple strategies have been considered for remedying random-effects misspecification. For example, Pinheiro et al. (2001) and Song et al. (2007) considered $t$ distributed random-effects and noise models, whereas Arellano-Valle et al. (2005), Lachos et al. (2010), and Ho and Lin (2010) considered the use of skew normal and $t$ distributed random and noise models. Although a rich class, the use of such distributions often do not allow for simplification of the marginal density (2) and do not allow for enough flexibility to model random-effects distributions with multiple modes or deviations from bell-shaped curves.

Based on the nonparametric maximum likelihood (NPML) principle of Laird (1978), Aitkin (1999) and Butler and Louis (1992) suggested using the point-mass density

$$f_{\boldsymbol{B}}(\boldsymbol{b}) = \sum_{i=1}^{g} \pi_i \delta(\boldsymbol{b} - \boldsymbol{\lambda}_i) \qquad (4)$$

for the random effects, where $\delta(\boldsymbol{x})$ is the Dirac delta function, $g \geq 1$ is the number of point masses, and $\boldsymbol{\lambda}_i \in \mathbb{R}^q$ and $\pi_k > 0$ are the point-mass locations and weights, for $i = 1, .., g$, respectively. To ensure that the total probability of the point masses adds up to unity and that the mean of $f_{\boldsymbol{B}}(\boldsymbol{b})$ is zero, we also require the restrictions that $\sum_{i=1}^{g} \pi_i = 1$ (which implies that $\pi_g = 1 - \sum_{i=1}^{g-1} \pi_i$) and $\sum_{i=1}^{g} \pi_i \boldsymbol{\lambda}_i = \boldsymbol{0}$, where $\boldsymbol{0}$ is a zero vector of appropriate size. We shall refer to densities of the form (4) as NPML-fitted (NPML) densities.

In Agresti et al. (2004) it was shown that NPML densities offered improvements in efficiency for estimating the characteristics of the random-effects density such as the covariance and individual *a posteriori* estimates of the random effects (i.e. $\mathbb{E}\left(\boldsymbol{B}_j | \boldsymbol{Y}_j = \boldsymbol{y}_j\right)$) when compared to the use of parametric random-effects densities in situations, where the true random-effect densities deviated from the assumed parametric form. However, this improvement comes at a cost of modeling complexity, since for any given $g$ and $q$, the number of parameters required for the specification and estimation of the NPML density is $(g-1)(q+1)$. This number can grow quickly if $g$ or $q$ are large.

We note that although unimodality or bell shape cannot be assumed in general, it may still be acceptable to assume symmetry in the random-effects distribution. Thus, we introduce the marginally symmetric NPML (MSMPML) random-effects density

$$f_{\boldsymbol{B}}\left(\boldsymbol{b}\right) = \sum_{i=1}^{g} \frac{\pi_i}{2}\left[\delta\left(\boldsymbol{b}-\boldsymbol{\lambda}_i\right) + \delta\left(\boldsymbol{b}+\boldsymbol{\lambda}_i\right)\right], \tag{5}$$

where $g$ and $\pi_i$ are as specified for (4) and no restrictions are made on $\boldsymbol{\lambda}_i$ as density (5) has zero mean by definition. For any $g$, density (5) comprises $2g$ point masses at $\boldsymbol{\lambda}_i$ and $-\boldsymbol{\lambda}_i$, each with weights $\pi_i/2$. Therefore, in situations where marginal symmetry can be assumed, the MSNPML model doubles the number of point masses when compared to the MPML density, for the same number of parameters.

Aside from the the aforementioned articles, our work shares commonalities with the mixture random-effects densities of Verbeke and Lesaffre (1996) and, by extension, the mixture of mixture of mixed effects models of Celeux et al. (2005) and Ng et al. (2006); in the aforementioned articles, mixtures of normal densities are used to model the random effects, rather than point-mass densities. When $n_j = N$ for all $j$, the MSNPML model corresponds to a symmetry-restricted and homoscedastic version of the repeated-measures linear mixture regression model of Grun and Leisch (2008). Furthermore, the use of parameter constraints to enforce marginal symmetry is similar to those used in Benaglia et al. (2009) and Chauveau and Hunter (2013). A kernel density approach to mixture modeling with symmetric components is presented in Chee and Wang (2013).

In this article, we shall consider the estimation of LMMs with the MSNPML random-effect density under the assumption of normal random error, using ML estimation via a multicycle expectation–conditional maximization (ECM) algorithm (Meng and Rubin, 1993). This algorithm is known to monotonically increase the log-likelihood and to lead to a stationary point of the said function in the case of convergence. Furthermore, we demonstrate how the covariance matrix and *a posteriori* expectations of the random effects can be estimated under model (5), and also consider various statistical matters such as consistency and the selection of the number of point masses. We then use numerical simulations to study the efficiency gains due to MSMPML over MPML, in the style of Agresti et al. (2004). A pair of example applications chosen from Pinheiro and Bates (2000) are subsequently used to demonstrate the inferential process of using MSMPML random effects.

The article shall proceed as follows. In Section 2, we shall discuss matters regarding ML estimation. In Section 3, statistical matters and computation of statistics are discussed. Results from simulations are then presented in Section 4, followed by example applications in Section 5. Conclusions are then drawn in Section 6.

## 2 Maximum Likelihood Estimation

Let $\boldsymbol{Y}_j$ be characterized by (1) and let $f_E(e)$ and $f_{\boldsymbol{B}}(\boldsymbol{b})$ have forms (3) and (5), respectively. Under this model, the individual marginal density (2) is given by

$$f_{\boldsymbol{Y}_j}(\boldsymbol{y}_j) = \sum_{i=1}^{g} \frac{\pi_i}{2} \left[ \prod_{k=1}^{n} \phi\left(y_{jk}; \boldsymbol{x}_{jk}^T \boldsymbol{\beta} + \boldsymbol{z}_{jk}^T \boldsymbol{\lambda}_i, \sigma^2\right) + \prod_{k=1}^{n} \phi\left(y_{jk}; \boldsymbol{x}_{jk}^T \boldsymbol{\beta} - \boldsymbol{z}_{jk}^T \boldsymbol{\lambda}_i, \sigma^2\right) \right] \tag{6}$$

and subsequently, the log-likelihood function has the form

$$l_n(\boldsymbol{\theta}; \boldsymbol{y}_1, ..., \boldsymbol{y}_n) = \sum_{j=1}^{n} \log f_{\boldsymbol{Y}_j}(\boldsymbol{y}_j), \tag{7}$$

where $\boldsymbol{\theta} = \left(\boldsymbol{\beta}^T, \boldsymbol{\lambda}^T, \boldsymbol{\pi}^T, \sigma^2\right)^T$ is the vector of all model parameters. Here $\boldsymbol{\lambda} = \left(\boldsymbol{\lambda}_1^T, ..., \boldsymbol{\lambda}_g^T\right)^T$, and $\boldsymbol{\pi} = (\pi_1, ..., \pi_{g-1})^T$. For brevity, we shall suppress the data dependence in $l_n(\boldsymbol{\theta}; \boldsymbol{y}_1, ..., \boldsymbol{y}_n)$ and write it as $l_n(\boldsymbol{\theta})$, where it causes no confusion.

In order to conduct ML estimation, one generally seeks an appropriate local maximizer of the log-likelihood function (7) (i.e. the ML estimator) over the parameter vector $\boldsymbol{\theta}$. However, due to the log-summation form of (7), it is not possible to obtain a maximizer in closed form.

We now present a multicycle ECM algorithm for the iterative computation of the ML estimator.

### 2.1 ECM Algorithm

By noting that the individual marginal densities (6) are each $2g$ component normal mixture densities, we can define the random variables $T_{(+i)j} = \mathbb{I}(\boldsymbol{b}_j = \boldsymbol{\lambda}_i)$ and $T_{(-i)j} = \mathbb{I}(\boldsymbol{b}_j = -\boldsymbol{\lambda}_i)$, and write the joint density of $\boldsymbol{Y}_j$ and $\boldsymbol{T}_j = \left(T_{(+1)j}, ..., T_{(+g)j}, T_{(-1)j}, ..., T_{(-g)j}\right)^T$

as

$$f_{T_j,Y_j}(t_j,y_j) = \prod_{i=1}^{g}\left(\left[\frac{\pi_i}{2}\prod_{k=1}^{n}\phi\left(y_{jk};x_{jk}^T\beta+z_{jk}^T\lambda_i,\sigma^2\right)\right]^{t_{(+i)j}}\right.$$
$$\left.\times\left[\frac{\pi_i}{2}\prod_{k=1}^{n}\phi\left(y_{jk};x_{jk}^T\beta-z_{jk}^T\lambda_i,\sigma^2\right)\right]^{t_{(-i)j}}\right),$$

where $\mathbb{I}(x = y)$ is an indicator variable that takes value 1 if $x = y$ and 0 otherwise. We call $f_{T_j,Y_j}(t_j,y_j)$ the individual complete-data density and use it to construct the complete-data log-likelihood in the form

$$\begin{aligned}c_n(\theta) &= \sum_{j=1}^{n}\log f_{T_j,Y_j}(t_j,y_j) \\ &= \sum_{i=1}^{g}\sum_{j=1}^{n}t_{(+i)j}\log\pi_i + \sum_{i=1}^{g}\sum_{j=1}^{n}t_{(-i)j}\log\pi_i - n\log 2 \\ &+ \sum_{i=1}^{g}\sum_{j=1}^{n}t_{(+i)j}\sum_{k=1}^{n_j}\log\phi\left(y_{jk};x_{jk}^T\beta+z_{jk}^T\lambda_i,\sigma^2\right) \\ &+ \sum_{i=1}^{g}\sum_{j=1}^{n}t_{(-i)j}\sum_{k=1}^{n_j}\log\phi\left(y_{jk};x_{jk}^T\beta-z_{jk}^T\lambda_i,\sigma^2\right).\end{aligned}$$

Since the ECM algorithm is iterative, we need to define some initial value $\theta^{(0)}$ and $m$th cycle update $\theta^{(m)}$. The $(m+1)$th cycle of the algorithm is conducted in two steps, the E-step and the CM-step.

In the $(m+1)$th E-step, we must compute the conditional expectation of $c_n(\theta)$ given the observed data and using the value of the parameter on the previous cycle, which can be expressed as

$$
\begin{aligned}
Q_n\left(\boldsymbol{\theta};\boldsymbol{\theta}^{(m)}\right) &= \mathbb{E}_{\boldsymbol{\theta}^{(m)}}\left[c_n\left(\boldsymbol{\theta}\right)|\boldsymbol{y}_1,...,\boldsymbol{y}_n\right] \\
&= \sum_{i=1}^{g}\sum_{j=1}^{n}\tau_{(+i)j}\left(\boldsymbol{\theta}^{(m)}\right)\log\pi_i + \sum_{i=1}^{g}\sum_{j=1}^{n}\tau_{(-i)j}\left(\boldsymbol{\theta}^{(m)}\right)\log\pi_i - n\log 2 \\
&\quad + \sum_{i=1}^{g}\sum_{j=1}^{n}\tau_{(+i)j}\left(\boldsymbol{\theta}^{(m)}\right)\sum_{k=1}^{n_j}\log\phi\left(y_{jk};\boldsymbol{x}_{jk}^T\boldsymbol{\beta}+\boldsymbol{z}_{jk}^T\boldsymbol{\rho}_i,\sigma^2\right) \\
&\quad + \sum_{i=1}^{g}\sum_{j=1}^{n}\tau_{(-i)j}\left(\boldsymbol{\theta}^{(m)}\right)\sum_{k=1}^{n_j}\log\phi\left(y_{jk};\boldsymbol{x}_{jk}^T\boldsymbol{\beta}-\boldsymbol{z}_{jk}^T\boldsymbol{\rho}_i,\sigma^2\right),
\end{aligned}
$$

where

$$
\begin{aligned}
\tau_{(+i)j}\left(\boldsymbol{\theta}\right) &= \mathbb{P}_{\boldsymbol{\theta}}\left(T_{(+i)j}|\boldsymbol{Y}_j=\boldsymbol{y}_j\right) \\
&= \frac{\pi_i\prod_{k=1}^{n_j}\phi\left(y_{jk};\boldsymbol{x}_{jk}^T\boldsymbol{\beta}+\boldsymbol{z}_{jk}^T\boldsymbol{\lambda}_i,\sigma^2\right)}{\sum_{i'=1}^{g}\pi_{i'}\left[\prod_{k=1}^{n_j}\phi\left(y_{jk};\boldsymbol{x}_{jk}^T\boldsymbol{\beta}+\boldsymbol{z}_{jk}^T\boldsymbol{\lambda}_{i'},\sigma^2\right)+\prod_{k=1}^{n_j}\phi\left(y_{jk};\boldsymbol{x}_{jk}^T\boldsymbol{\beta}-\boldsymbol{z}_{jk}^T\boldsymbol{\lambda}_{i'},\sigma^2\right)\right]}
\end{aligned} \quad (8)
$$

and

$$
\begin{aligned}
\tau_{(-i)j}\left(\boldsymbol{\theta}\right) &= \mathbb{P}_{\boldsymbol{\theta}}\left(T_{(-i)j}|\boldsymbol{Y}_j=\boldsymbol{y}_j\right) \\
&= \frac{\pi_i\prod_{k=1}^{n_j}\phi\left(y_{jk};\boldsymbol{x}_{jk}^T\boldsymbol{\beta}-\boldsymbol{z}_{jk}^T\boldsymbol{\lambda}_i,\sigma^2\right)}{\sum_{i'=1}^{g}\pi_{i'}\left[\prod_{k=1}^{n_j}\phi\left(y_{jk};\boldsymbol{x}_{jk}^T\boldsymbol{\beta}+\boldsymbol{z}_{jk}^T\boldsymbol{\lambda}_{i'},\sigma^2\right)+\prod_{k=1}^{n_j}\phi\left(y_{jk};\boldsymbol{x}_{jk}^T\boldsymbol{\beta}-\boldsymbol{z}_{jk}^T\boldsymbol{\lambda}_{i'},\sigma^2\right)\right]}.
\end{aligned} \quad (9)
$$

In the $(m+1)$th CM-step, we seek to obtain the maximizer of $Q_n\left(\boldsymbol{\theta};\boldsymbol{\theta}^{(m)}\right)$ over the space of $\boldsymbol{\theta}$. However, due to the form of $Q_n\left(\boldsymbol{\theta};\boldsymbol{\theta}^{(m)}\right)$, it is simpler to maximize over the parameter space holding either $\boldsymbol{\beta}$ or $\boldsymbol{\lambda}$ fixed in each cycle; this partial update over the parameter vector is the difference between the EM and ECM algorithms (cf. McLachlan and Krishnan (2008, Sec. 5.2)).

Without loss of generality, if $m$ is even, the $(m+1)$th CM-step is conducted by setting $\boldsymbol{\lambda}^{(m+1)} = \boldsymbol{\lambda}^{(m)}$ and computing the update

$$\boldsymbol{\beta}^{(m+1)} = \left( \sum_{j=1}^{n} \sum_{k=1}^{n_j} \boldsymbol{x}_{jk} \boldsymbol{x}_{jk}^T \right)^{-1} \boldsymbol{u}^{(m)}, \qquad (10)$$

where

$$\begin{aligned}
\boldsymbol{u}^{(m)} &= \sum_{i=1}^{g} \sum_{j=1}^{n} \tau_{(+i)j}\left(\boldsymbol{\theta}^{(m)}\right) \sum_{k=1}^{n_j} \boldsymbol{x}_{jk} \left( y_{jk} - \boldsymbol{z}_{jk}^T \boldsymbol{\lambda}_i^{(m)} \right) \\
&+ \sum_{i=1}^{g} \sum_{j=1}^{n} \tau_{(-i)j}\left(\boldsymbol{\theta}^{(m)}\right) \sum_{k=1}^{n_j} \boldsymbol{x}_{jk} \left( y_{jk} + \boldsymbol{z}_{jk}^T \boldsymbol{\lambda}_i^{(m)} \right).
\end{aligned}$$

If $m$ is odd, we instead set $\boldsymbol{\beta}^{(m+1)} = \boldsymbol{\beta}^{(m)}$ and compute the updates

$$\boldsymbol{\lambda}_i^{(m+1)} = \left( \sum_{j=1}^{n} \sum_{k=1}^{n_j} \boldsymbol{z}_{jk} \boldsymbol{z}_{jk}^T \right)^{-1} \boldsymbol{v}_i^{(m)} \qquad (11)$$

for each $i$, where

$$\begin{aligned}
\boldsymbol{v}_i^{(m)} &= \sum_{j=1}^{n} \tau_{(+i)j}\left(\boldsymbol{\theta}^{(m)}\right) \sum_{k=1}^{n_j} \boldsymbol{z}_{jk} \left( y_{jk} - \boldsymbol{x}_{jk}^T \boldsymbol{\beta}^{(m)} \right) \\
&- \sum_{j=1}^{n} \tau_{(-i)j}\left(\boldsymbol{\theta}^{(m)}\right) \sum_{k=1}^{n_j} \boldsymbol{z}_{jk} \left( y_{jk} - \boldsymbol{x}_{jk}^T \boldsymbol{\beta}^{(m)} \right).
\end{aligned}$$

Lastly, in all CM-steps, we update the remaining parameter components by computing

$$\pi_i^{(m+1)} = \frac{\sum_{j=1}^{n} \tau_{(-i)j}\left(\boldsymbol{\theta}^{(m)}\right) + \sum_{j=1}^{m} \tau_{(+i)j}\left(\boldsymbol{\theta}^{(m)}\right)}{n} \qquad (12)$$

for each $i$, and

$$\sigma^{2(m+1)} = \frac{w^{(m+1)}}{\sum_{j=1}^{n} n_j}, \qquad (13)$$

where

$$w^{(m+1)} = \sum_{i=1}^{g}\sum_{j=1}^{n}\tau_{(+i)j}\left(\boldsymbol{\theta}^{(m)}\right)\sum_{k=1}^{n_j}\left(y_{jk}-\boldsymbol{x}_{jk}^T\boldsymbol{\beta}^{(m+1)}-\boldsymbol{z}_{jk}^T\boldsymbol{\lambda}_i^{(m+1)}\right)^2$$
$$+\sum_{i=1}^{g}\sum_{j=1}^{n}\tau_{(-i)j}\left(\boldsymbol{\theta}^{(m)}\right)\sum_{k=1}^{n_j}\left(y_{jk}-\boldsymbol{x}_{jk}^T\boldsymbol{\beta}^{(m+1)}+\boldsymbol{z}_{jk}^T\boldsymbol{\lambda}_i^{(m+1)}\right)^2.$$

We note that an iteration of the ECM algorithm is characterized by the updating of all parameter components, and is thus made up of two consecutive cycles.

## 2.2 Convergence Analysis

The E- and CM-steps are iterated until a convergence criterion is met, at which stage we declare the final iterate the ML estimate $\hat{\boldsymbol{\theta}}_n$. In this article, we choose to use the absolute convergence criterion

$$l_n\left(\boldsymbol{\theta}^{(m+1)}\right)-l_n\left(\boldsymbol{\theta}^{(m)}\right)<\epsilon$$

for some small tolerance $\epsilon>0$; for all applications, we set $\epsilon=10^{-8}$. See Lange (2013, Sec. 11.5) for a discussion on the relative merits of convergence criteria for optimization algorithms.

Let $\boldsymbol{\theta}^*=\lim_{m\to\infty}\boldsymbol{\theta}^{(m)}$ (alternatively $\hat{\boldsymbol{\theta}}_n\to\boldsymbol{\theta}^*$ as $\epsilon\to 0$) be a limit point of a sequence $\boldsymbol{\theta}^{(m)}$ of the ECM algorithm, for some starting value $\boldsymbol{\theta}_0$. In order to infer the properties of $\boldsymbol{\theta}^*$ and the ECM algorithm, we make the following observations.

Firstly, consider that for even and odd $m$, the updates uniquely satisfy the

equations

$$\left(\boldsymbol{\beta}^{(m+1)T}, \boldsymbol{\pi}^{(m+1)T}, \sigma^{2(m+1)}\right)^T = \arg\max_{\boldsymbol{\beta}, \boldsymbol{\pi}, \sigma^2} Q_n\left(\boldsymbol{\beta}, \boldsymbol{\lambda}^{(m)}, \boldsymbol{\pi}, \sigma^2; \boldsymbol{\theta}^{(m)}\right) + \eta\left(\boldsymbol{\theta}^{(m)}\right) \tag{14}$$

and

$$\left(\boldsymbol{\lambda}^{(m+1)T}, \boldsymbol{\pi}^{(m+1)T}, \sigma^{2(m+1)}\right)^T = \arg\max_{\boldsymbol{\lambda}, \boldsymbol{\pi}, \sigma^2} Q_n\left(\boldsymbol{\beta}^{(m)}, \boldsymbol{\lambda}, \boldsymbol{\pi}, \sigma^2; \boldsymbol{\theta}^{(m)}\right) + \eta\left(\boldsymbol{\theta}^{(m)}\right) \tag{15}$$

respectively, over the 2active parameter subsets in each problem (i.e. either holding $\boldsymbol{\beta}$ or $\boldsymbol{\lambda}$ constant). Here

$$\eta\left(\boldsymbol{\theta}\right) = -\sum_{i=1}^{g}\sum_{j=1}^{n}\left[\tau_{(-i)j}\left(\boldsymbol{\theta}\right)\log\tau_{(-i)j}\left(\boldsymbol{\theta}\right) + \tau_{(+i)j}\left(\boldsymbol{\theta}\right)\log\tau_{(+i)j}\left(\boldsymbol{\theta}\right)\right]$$

is known as the entropy; see McLachlan and Peel (2000, Sec. 6.10).

Next, we note that the functions from (14) and (15) are both quasi-convex and satisfy Assumption 2 of Razaviyayn et al. (2013) and are thus both lower-bound functions of (7), over the respective parameter subsets (minorizers in the language of minorization–maximization algorithms; see Lange (2013, Ch. 8)). Finally, we observe that (7) is everywhere smooth; taken all together, this implies that the ECM algorithm satisfies the conditions for Corollary 2 of Razaviyayn et al. (2013), which yields the following convergence result.

**Proposition 1.** *Let $\boldsymbol{\theta}^{(m)}$ be a sequence of ECM algorithm iterates with limit point $\boldsymbol{\theta}^*$, for some initial value $\boldsymbol{\theta}^{(0)}$. If $\boldsymbol{\theta}^{(m+1)}$ is updated via the steps $\boldsymbol{\lambda}^{(m+1)} = \boldsymbol{\lambda}^{(m)}$, (10), (12), and (13), when m is even; and $\boldsymbol{\beta}^{(m+1)} = \boldsymbol{\beta}^{(m)}$, (11), (12), and (13), when m is odd, then the following statements are true.*

*1. The sequence of log-likelihoods $l_n\left(\boldsymbol{\theta}^{(m)}\right)$ is monotonically increasing in m.*

*2. The limit point $\boldsymbol{\theta}^*$ is a stationary point of $l_n\left(\boldsymbol{\theta}\right)$.*

Thus, Proposition 1 implies that the ECM algorithm presented exhibits the usual monotonicity property of EM-type algorithms, and that the obtained ML estimate is convergent to a stationary point of the log-likelihood function. We note that the log-likelihood functions of mixture models tend to be highly multimodal. Thus care is required when optimizing such functions as different starting values $\boldsymbol{\theta}^{(0)}$ may lead to different optima or saddle-points. Throughout this article, we use multiple random starting values (McLachlan and Peel, 2000, Sec. 2.12) to mitigate against such problems in order to obtain good ML estimates. Finally, we remark that we have chosen to check the Razaviyayn et al. (2013) conditions to establish the convergence properties of the algorithm, due to their simplicity; however, it is also possible to obtain the convergence results via the conditions of Meng and Rubin (1993).

## 3 Statistical Inference

### 3.1 Consistency and Asymptotic Normality

Establishing consistency and asymptotic normality are fundamental to the process of drawing accurate inferences with the parameter estimate. In the case where all $n_j = N$ for some constant $N \geq 2$ (i.e. the balanced experiment setting, in the language of LMMs) we can observe that the density function (6) is simply that of a $2g$ component $N$-dimensional mixture of multivariate normal distributions. Thus, in this setting, consistency and asymptotic normality can be established via conventional means for establishing the asymptotic properties of ML estimators, such as by Theorem 4.2.4 of Amemiya (1985). The validity of the assumptions for such theorems can be assessed via the techniques of Atienza et al. (2007). As a consequence, Theorem 4.2.4 of Amemiya (1985) establishes the following result regarding the ML estimation of the MSMPML LMM.

**Proposition 2.** *Assume that the data $Y_1, ..., Y_n$ is an IID sample that arises from an MSMPML LMM with parameter $\boldsymbol{\theta}_0$, where the covariates $\boldsymbol{x}_{jk}$ and $\boldsymbol{z}_{jk}$ are realizations from a well-behaved random process, for all $j$ and $k$. Let $\boldsymbol{\Theta}_n$ be the set of roots of the equation $\partial l_n(\boldsymbol{\theta})/\partial \boldsymbol{\theta} = \mathbf{0}$ that corresponds to the local maxima of $l_n(\boldsymbol{\theta})$ (here $l_n(\boldsymbol{\theta})$ is as in (7) and $\boldsymbol{\Theta}_n = \{\mathbf{0}\}$ if there are no roots corresponding to a maximum). If $n_j = N$ for all $j$, where $N \geq 2$, then the following results hold.*

1. *For any $\epsilon > 0$,*

$$\lim_{n \to \infty} \mathbb{P}\left[\inf_{\boldsymbol{\theta} \in \boldsymbol{\Theta}_n} (\boldsymbol{\theta} - \boldsymbol{\theta}_0)^T (\boldsymbol{\theta} - \boldsymbol{\theta}_0) > \epsilon\right] = 0.$$

2. *If $\hat{\boldsymbol{\theta}}_n \xrightarrow{P} \boldsymbol{\theta}_0$, then*

$$\sqrt{n}\left(\hat{\boldsymbol{\theta}}_n - \boldsymbol{\theta}_0\right) \xrightarrow{L} \mathcal{N}\left(\mathbf{0}, \boldsymbol{I}^{-1}(\boldsymbol{\theta}_0)\right),$$

*where*

$$\boldsymbol{I}(\boldsymbol{\theta}) = \mathbb{E}\left[-\frac{1}{n}\frac{\partial^2 l_n(\boldsymbol{\theta}; \boldsymbol{Y}_1, ..., \boldsymbol{Y}_n)}{\partial \boldsymbol{\theta} \partial \boldsymbol{\theta}^T}\right]$$

*is the Fisher information matrix.*

We make the following notes regarding the application of Proposition 2. Firstly, the assumption that $\boldsymbol{x}_{jk}$ and $\boldsymbol{z}_{jk}$ be well-behaved is left purposefully vague. This assumption simply stipulates that the covariates be such that the uniform convergence of the average log-likelihood and its derivatives hold. For example, one can assume that $\boldsymbol{X}_{jk}$ and $\boldsymbol{Z}_{jk}$ have a continuous joint density over a compact support, and are independent for all $i$ and $j$.

Next, Part 1 of the proposition simply states that there exists a consistent root of the log-likelihood (7). As noted in Section 2.2, the log-likelihood is likely to be multimodal. Thus, one needs to choose carefully, which of the roots to declare as the ML estimate. We acknowledge that the existence of a consistent root allows for bypassing of the issue of identifiability, as the proposition establishes that at least one of the permutations of possible solutions is a consistent estimate. We make a further note that although Part 2 establishes asymptotic normality, due to identifiability issues, some hypothesis test statistics do not retain their usual asymptotic distributions; see Boos and Stefanski (2013, Sec. 6.6.5) and McLachlan and Peel (2000, Sec. 6.5) for discussions on such issues.

Lastly, to extend the consistency and asymptotic normality of the ML estimator to the setting of non-random covariates, one simply is required to make extra assumptions regarding the uniform convergence of the average log-likelihood and its derivatives; see Theorem 4.2.4 of Amemiya (1985) for details. However, it is non-trivial and beyond the scope of this article to extend Proposition 2 to the unbalanced setting (i.e. when $n_j$ are allowed to be different, for all $j$). We direct the interested reader to Miller (1977) and Weiss (1971) for treatments on the asymptotic of ML estimators in such settings.

As noted by a reviewer, we have not discussed the identifiability of models of form (2). Since we are only using the model in order to nonparametrically infer the properties of the random-effects density function (e.g. the covariance matrix and the shape), we do not require (2) to be identifiable in the sense of Titterington et al. (1985, Sec. 3.1). However, it is desirable to establish a notion of asymptotic identifiability, as considered by Redner (1981). Because (2) can be viewed as a $2g$-component mixture model, the results of Atienza et al. (2007) establishes such an asymptotic identifiability result.

### 3.2 Estimation of Statistics

As noted in Section 1, in LMMs, it is interesting to quantify the covariance of the random-effects distribution, as well as individual *a posteriori* estimates of random effects. Firstly, since the MSNPML random-effects distribution 5 is a point-mass distribution with mean $\mathbf{0}$, the definition of a covariance matrix

$$\text{cov}\,(\boldsymbol{B}) = \mathbb{E}\left([\boldsymbol{B} - \mathbb{E}\boldsymbol{B}][\boldsymbol{B} - \mathbb{E}\boldsymbol{B}]^T\right)$$

yields the simple expression,

$$\text{cov}\,(\boldsymbol{B}) = \sum_{i=1}^{g} \pi_i \boldsymbol{\lambda}_i \boldsymbol{\lambda}_i^T. \tag{16}$$

Secondly, using the a posteriori distribution for each individual $j$, as characterized by (8) and (9) (which is also a point-mass distribution), we can write the *a posteriori* expectation for each individual as

$$\mathbb{E}\,(\boldsymbol{B}_j|\boldsymbol{Y}_j = \boldsymbol{y}_j) \;\;=\;\; \frac{\sum_{i=1}^{g}\left[\tau_{(+i)j}(\boldsymbol{\theta}) - \tau_{(-i)j}(\boldsymbol{\theta})\right]\boldsymbol{\lambda}_i}{\sum_{i=1}^{g}\left[\tau_{(+i)j}(\boldsymbol{\theta}) + \tau_{(-i)j}(\boldsymbol{\theta})\right]}. \tag{17}$$

Of course, we do not have access to the true parameter vector $\boldsymbol{\theta}_0$, in order to evaluate the quantities above at their true values. However, as a consequence of Proposition 2 and continuous mapping, replacing $\boldsymbol{\theta}$ elements in either (16) or (17) with elements from $\hat{\boldsymbol{\theta}}_n$ will yield consistent estimates of the respective quantities; we shall denote these estimates by $\hat{\boldsymbol{V}}$ and $\hat{\boldsymbol{b}}_j$, respectively.

### 3.3 Number of Point Masses

In the discussion thus far, it has been assumed that the number of point masses $2g$ is fixed. This is due to the semi-parametric formulation of the MSNPML

LMM, which makes it is difficult to incorporate the determination of $g$ within the maximization process described in Section 2.

Due to the mixture model form of the marginal density (6), we can utilize the techniques available from the mixture model literature, for the determination of the number of components in a model; see McLachlan and Peel (2000, Ch. 6). Much of the literature on hypothesis testing for determining $g$ is bespoken to specific distributions and contexts, or requires computationally intensive resampling techniques; however, the information theoretical approach to components selection remains relatively portable. Thus, in this article, we shall use the Akaike information criterion (AIC) (Akaike, 1974) and Bayesian information criterion (BIC) (Schwarz, 1978) for the purpose of determining $g$.

Let $G = \{g_1, ..., g_M\}$ be a set of $M$ potential values for $g$, where $g_h \geq 1$ for each $h = 1, ..., M$. Furthermore, let $g_0 \in G$, where $g_0$ is the true value of $g$. For each $h$, we fit the MSNPML LMM with $2g_h$ point masses via the ECM algorithm to obtain the ML estimate $\hat{\boldsymbol{\theta}}_{(h)n}$. Using the estimates, we can then compute the AIC and BIC values for each model as

$$\text{AIC}(h) = -2l_n\left(\hat{\boldsymbol{\theta}}_{(h)n}\right) + 2\left[g_h(q+1) + p\right]$$

and

$$\text{BIC}(h) = -2l_n\left(\hat{\boldsymbol{\theta}}_{(h)n}\right) + \log\left(\sum_{j=1}^{n} n_j\right)\left[g_h(q+1) + p\right],$$

respectively. Here, $g_h(q+1) + p$ is the total number of parameter elements in the model with $2g_h$ point masses.

Using either of the criteria, the rule is to choose to set $g = g_h$, for which $h = 1, ..., M$ is the argument which minimizes the criterion in use. If the assumption that $g_0 \in G$ is valid, then under some regularity conditions, it is provable that the AIC rule will asymptotically select a model where $g \geq g_0$ (cf. Leroux (1992,

Sec. 3)) and that the BIC rule will asymptotically select a model where $g = g_0$ (cf. Keribin (2000, Sec. 3)). Simulation results in McLachlan and Peel (2000, Sec. 6.11) show that both criteria are capable of selecting the correct number of components in simple mixture model situations, and as expected from the theory, the BIC rule tends to outperform the AIC rule. Furthermore, results from Grun and Leisch (2007), Depraetere and Vandebroek (2014), and Nguyen and McLachlan (2016) confirm this observation in regression settings.

## 4 Numerical Simulations

We now perform a set of numerical simulation studies to assess the performance of MSNPML versus the NPML random-effects model, when the random-effects density is marginally symmetric. Our simulations follows in the style of Agresti et al. (2004) and are conducted as follows.

All computation and random data generation in Sections 4 and 5 are conducted within the $R$ computing environment (R Core Team, 2013). Data generation of exponential, normal, and uniform random variates are conducted using the *rexp*, *rnorm*, and *runif* functions of the *stats* package (R Core Team, 2013), respectively. Data generation of triangular random variates are conducted using the *rtriangle* function from the *triangle* package (Carnell, 2013). The ECM algorithms for both the NPML and MSNPML models are programmed in $R$, and computationally intensive components are implemented in $C$ and integrated into $R$ using the *Rcpp* and *RcppArmadillo* packages (Eddelbuettel, 2013).

### 4.1 Simulation Setup

We simulate $R = 1000$ instances of $n \in \{50, 100\}$ individuals, each with $n_j = N \in \{50, 100\}$ responses according to model (1), where $\boldsymbol{\beta} = (0,0)^T$, $\sigma^2 = 1$, and $\boldsymbol{x}_{jk} = \boldsymbol{z}_{jk} = (1, \nu_{jk})^T$, where $\nu_{jk}$ is a realization from Uniform $(-1, 1)$, for each $i$ and $j$. The random effects $\boldsymbol{b}_j = (b_{1j}, b_{2j})^T$ are simulated under the following six mean $\boldsymbol{0}$ scenarios: uniform over $(-1,1)^2$ (S1); multivariate normal with mean $\boldsymbol{0}$ and covariance $\text{diag}(1,1)$ (S2); equal mixture of two uniform distributions over $(-2, 0) \times (0, 2)$ and $(0, 2) \times (-2, 0)$ (S3); equal mixture of two multivariate normal distributions with means and covariance matrices $(1, -1)^T$ and $\text{diag}(1/4, 1/4)$, and $(-1, 1)^T$ and $\text{diag}(1/4, 1/4)$ (S4); $\boldsymbol{B}_j = (C_1 - 1/3, C_2 - 1/3)^T$, where $C_1$ and $C_2$ are independent and triangularly distributed with parameters $(-1, 0, 2)^T$ (cf. Kotz and Van Dorp (2004, Ch. 1)) (S5); and $\boldsymbol{B}_j = (D_1 - 1/12, D_2 - 1/12)^T$, were $D_1$ and $D_2$ are independently observations from an equal mixture of an exponential distribution with mean $1/2$ and a negative exponential distribution with mean $-1/3$ (S6). This yields a total of 20 different simulation scenarios. Here, S1–S4 are marginally symmetric whereas S5 and S6 are not. Figure 1 shows a single instance of $n = 100$ realizations of the random-effects vectors $\boldsymbol{B}_j$ under each of the six scenarios.

We can write the covariance matrix of the random-effects distribution as

$$\text{cov}(\boldsymbol{B}) = \begin{bmatrix} V_{11} & V_{12} \\ V_{12} & V_{22} \end{bmatrix}, \qquad (18)$$

where, we note that $V_{11} = V_{22}$ in all six scenarios. The exact values of $V_{11}$ and $V_{12}$ for each scenario are as follows: $V_{11} = 1/3$ and $V_{12} = 0$ in S1; $V_{11} = 1$ and $V_{12} = 0$ in S2; $V_{11} = 4/3$ and $V_{12} = -1$ in S3; $V_{11} = 5/4$ and $V_{12} = -1$ in S4; $V_{11} = 7/18$ and $V_{12} = 0$ in S5; and $V_{11} = 17/48$ and $V_{12} = 0$ in S6.

For each $r = 1, ..., R$, and under each of the 20 scenarios, we compute an ML estimate $\hat{\boldsymbol{\theta}}_{[r]}$ for both the MSNPML and NPML LMMs with $g \in \{2, 3, 4, 5\}$ (here, we skip $g = 1$ because the NPML model would have no variability). Using the same approach as Agresti et al. (2004), we then compute the efficiency of each set of estimates via the mean absolute deviation (MAD) criterion. That is, to assess the fixed-effects estimates, we compute $R^{-1}\sum_{r=1}^{R}\left|\hat{\beta}_{[r]h} - \beta_h\right|$ for $h = 1, 2$; to assess the error variance estimates, we compute $R^{-1}\sum_{r=1}^{R}\left|\hat{\sigma}_{[r]}^2 - \sigma^2\right|$; and to assess the covariance matrix estimate, we compute $R^{-1}\sum_{r=1}^{R}\left|\hat{V}_{[r]h_1h_2} - V_{l_1l_2}\right|$ for $h_1 \leq h_2$. Here

$$\hat{\boldsymbol{V}}_{[r]} = \begin{bmatrix} \hat{V}_{[r]11} & \hat{V}_{[r]12} \\ \hat{V}_{[r]12} & \hat{V}_{[r]22} \end{bmatrix}$$

is the estimated covariance matrix using the ML estimate $\hat{\boldsymbol{\theta}}_{[r]}$.

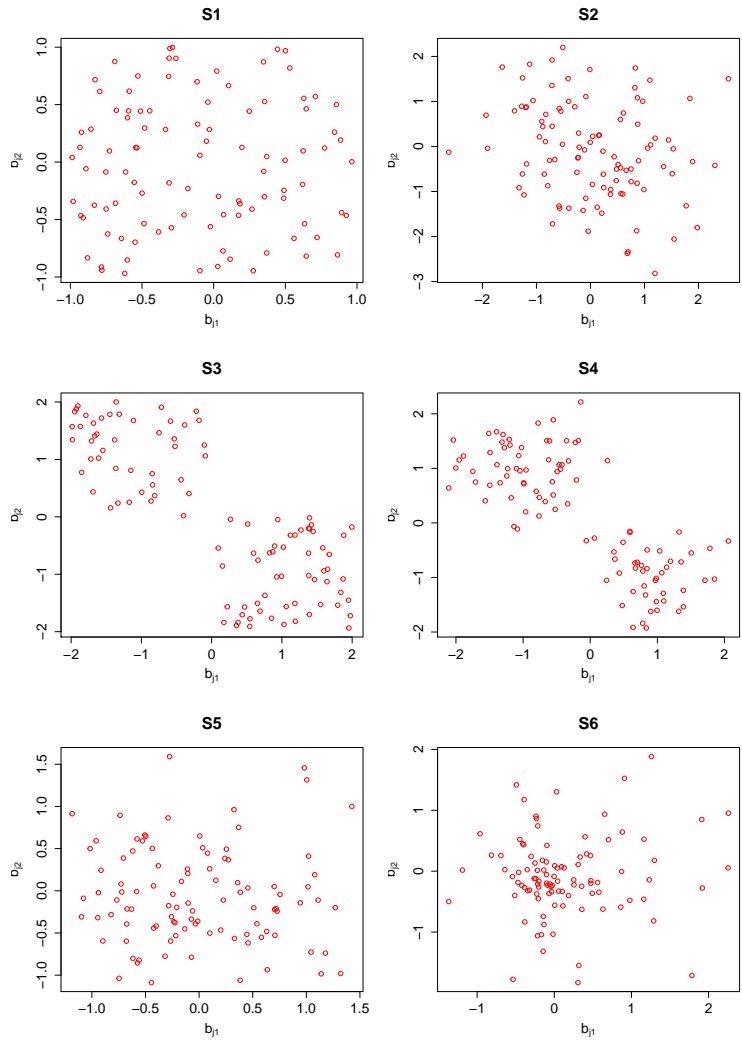

Figure 1: Typical samples of $n = 100$ realizations from each of the four simulation scenarios described in Section 4.1.

Further, for each $r$, under each of the 20 scenarios, we compute the average AIC and BIC values for both the MSNPML and NPML LMMs with $g \in \{2, 3, 4, 5\}$, and determine the number of components $g$ that minimize the average values over the 20 scenarios. Here, we also compare the MAD of the covariance matrix estimates between the most selected number of components of the two models.

## 4.2 Results

The results of simulations S1–S4 are reported in Tables 1–4, respectively. Firstly, we make the general observation regarding the MADs in each of these scenarios. The MSNPML model appears to uniformly outperform the NPML model with respect to the estimation of the marginal variances $V_{11}$ and $V_{22}$, and the error variance $\sigma^2$. Here, on many occasions, the MAD of the MSNPML model estimates are approximately 1/3 that of the corresponding NPML model estimates.

Table 1: Simulation results for S1. Italics indicate models with the lowest AIC values, and bold texts indicate models with the lowest BIC values. Columns $\beta_1$, $\beta_2$, $\sigma^2$, $V_{11}$, $V_{12}$, and $V_{13}$ present MAD values of the respective parameter elements, for each model.

| $n$ | $n_j = N$ | Model | $g$ | AIC | BIC | $\beta_1$ | $\beta_2$ | $\sigma^2$ | $V_{11}$ | $V_{12}$ | $V_{22}$ |
|---|---|---|---|---|---|---|---|---|---|---|---|
| 50 | 50 | MSNPML | 2 | 7411.84 | 7458.43 | 0.0740 | 0.0811 | 0.0722 | 0.0527 | 0.0484 | 0.1120 |
| | | | 3 | 7353.49 | 7417.56 | 0.0695 | 0.0807 | 0.0348 | 0.0403 | 0.0453 | 0.0640 |
| | | | 4 | 7337.43 | **7418.97** | **0.0673** | **0.0830** | **0.0252** | **0.0397** | **0.0450** | **0.0520** |
| | | | 5 | *7333.31* | 7432.32 | *0.0669* | *0.0852* | *0.0231* | *0.0390* | *0.0445* | *0.0495* |
| | | NPML | 2 | 7571.67 | 7618.26 | 0.0660 | 0.0745 | 0.1720 | 0.0767 | 0.0479 | 0.3190 |
| | | | 3 | 7475.32 | 7539.39 | 0.0658 | 0.0740 | 0.1090 | 0.0590 | 0.0469 | 0.2110 |
| | | | 4 | 7413.81 | **7495.34** | **0.0658** | **0.0738** | **0.0686** | **0.0508** | **0.0464** | **0.1260** |
| | | | 5 | *7378.88* | 7477.89 | *0.0658* | *0.0734* | *0.0449* | *0.0449* | *0.0436* | *0.0850* |
| 50 | 100 | MSNPML | 2 | 14844.99 | 14897.13 | 0.0670 | 0.0754 | 0.0857 | 0.0544 | 0.0455 | 0.1290 |
| | | | 3 | 14715.44 | 14787.13 | 0.0636 | 0.0752 | 0.0446 | 0.0407 | 0.0421 | 0.0735 |
| | | | 4 | 14679.03 | **14770.27** | **0.0596** | **0.0781** | **0.0284** | **0.0388** | **0.0411** | **0.0538** |
| | | | 5 | *14664.63* | 14775.42 | *0.0601* | *0.0791* | *0.0213* | *0.0376* | *0.0415* | *0.0473* |
| | | NPML | 2 | 15150.45 | 15202.58 | 0.0667 | 0.0683 | 0.1790 | 0.0817 | 0.0457 | 0.3200 |
| | | | 3 | 14973.52 | 15045.21 | 0.0666 | 0.0682 | 0.1210 | 0.0617 | 0.0438 | 0.2270 |
| | | | 4 | 14845.31 | 14936.55 | 0.0665 | 0.0680 | 0.0824 | 0.0531 | 0.0427 | 0.1440 |
| | | | 5 | *14763.89* | **14874.68** | **0.0666** | **0.0681** | **0.0581** | **0.0466** | **0.0413** | **0.1000** |
| 100 | 50 | MSNPML | 2 | 14744.46 | 14796.59 | 0.0513 | 0.0548 | 0.0813 | 0.0486 | 0.0387 | 0.1120 |
| | | | 3 | 14594.81 | 14666.50 | 0.0481 | 0.0512 | 0.0379 | 0.0316 | 0.0321 | 0.0583 |
| | | | 4 | 14542.16 | 14633.40 | 0.0475 | 0.0556 | 0.0237 | 0.0292 | 0.0315 | 0.0436 |
| | | | 5 | *14517.74* | **14628.53** | **0.0479** | **0.0570** | **0.0184** | **0.0284** | **0.0316** | **0.0358** |
| | | NPML | 2 | 15091.32 | 15143.46 | 0.0485 | 0.0499 | 0.1790 | 0.0747 | 0.0347 | 0.3260 |
| | | | 3 | 14877.60 | 14949.29 | 0.0485 | 0.0496 | 0.1200 | 0.0545 | 0.0347 | 0.2270 |
| | | | 4 | 14737.22 | 14828.46 | 0.0484 | 0.0496 | 0.0792 | 0.0462 | 0.0360 | 0.1290 |
| | | | 5 | *14648.40* | **14759.19** | **0.0485** | **0.0495** | **0.0522** | **0.0379** | **0.0318** | **0.0800** |
| 100 | 100 | MSNPML | 2 | 29536.47 | 29594.15 | 0.0502 | 0.0524 | 0.0914 | 0.0523 | 0.0342 | 0.1300 |
| | | | 3 | 29215.19 | 29294.51 | 0.0464 | 0.0525 | 0.0479 | 0.0333 | 0.0290 | 0.0703 |
| | | | 4 | 29104.46 | 29205.40 | 0.0425 | 0.0524 | 0.0311 | 0.0293 | 0.0283 | 0.0543 |
| | | | 5 | *29049.10* | **29171.68** | **0.0435** | **0.0550** | **0.0220** | **0.0283** | **0.0280** | **0.0408** |
| | | NPML | 2 | 30215.66 | 30273.35 | 0.0478 | 0.0490 | 0.1840 | 0.0809 | 0.0317 | 0.3270 |
| | | | 3 | 29822.61 | 29901.93 | 0.0477 | 0.0489 | 0.1290 | 0.0573 | 0.0315 | 0.2410 |
| | | | 4 | 29524.39 | 29625.33 | 0.0478 | 0.0490 | 0.0894 | 0.0491 | 0.0321 | 0.1480 |
| | | | 5 | *29334.68* | **29457.26** | **0.0478** | **0.0488** | **0.0632** | **0.0412** | **0.0285** | **0.0973** |

Table 2: Simulation results for S2. Italics indicate models with the lowest AIC values, and bold texts indicate models with the lowest BIC values. Columns $\beta_1$, $\beta_2$, $\sigma^2$, $V_{11}$, $V_{12}$, and $V_{13}$ present MAD values of the respective parameter elements, for each model.

| $n$ | $n_j = N$ | Model | $g$ | AIC | BIC | $\beta_1$ | $\beta_2$ | $\sigma^2$ | $V_{11}$ | $V_{12}$ | $V_{22}$ |
|---|---|---|---|---|---|---|---|---|---|---|---|
| 50 | 50 | MSNPML | 2 | 7942.22 | 7988.81 | 0.1780 | 0.1490 | 0.3250 | 0.2020 | 0.1380 | 0.5120 |
| | | | 3 | 7743.67 | 7807.74 | 0.1880 | 0.1680 | 0.2020 | 0.1730 | 0.1320 | 0.3350 |
| | | | 4 | 7642.53 | 7724.07 | 0.1820 | 0.1850 | 0.1390 | 0.1700 | 0.1280 | 0.2450 |
| | | | 5 | *7577.92* | **7676.93** | ***0.1850*** | ***0.1850*** | ***0.0978*** | ***0.1660*** | ***0.1270*** | ***0.2070*** |
| | | NPML | 2 | 8405.01 | 8451.61 | 0.1130 | 0.1180 | 0.6390 | 0.3410 | 0.1390 | 0.9490 |
| | | | 3 | 8118.61 | 8182.68 | 0.1130 | 0.1170 | 0.4370 | 0.2360 | 0.1380 | 0.7500 |
| | | | 4 | 7928.83 | 8010.37 | 0.1130 | 0.1170 | 0.3130 | 0.2070 | 0.1260 | 0.5060 |
| | | | 5 | *7808.73* | **7907.74** | ***0.1130*** | ***0.1170*** | ***0.2390*** | ***0.1880*** | ***0.1210*** | ***0.3900*** |
| 50 | 100 | MSNPML | 2 | 15965.79 | 16017.93 | 0.1780 | 0.1470 | 0.3410 | 0.2100 | 0.1270 | 0.5330 |
| | | | 3 | 15566.11 | 15637.80 | 0.1770 | 0.1610 | 0.2190 | 0.1780 | 0.1220 | 0.3590 |
| | | | 4 | 15360.97 | 15452.21 | 0.1760 | 0.1670 | 0.1590 | 0.1690 | 0.1150 | 0.2610 |
| | | | 5 | *15218.55* | **15329.34** | ***0.1770*** | ***0.1750*** | ***0.1180*** | ***0.1640*** | ***0.1160*** | ***0.2220*** |
| | | NPML | 2 | 16859.61 | 16911.75 | 0.1150 | 0.1160 | 0.6430 | 0.3550 | 0.1270 | 0.9560 |
| | | | 3 | 16323.08 | 16394.77 | 0.1150 | 0.1170 | 0.4490 | 0.2550 | 0.1230 | 0.7460 |
| | | | 4 | 15948.04 | 16039.28 | 0.1150 | 0.1170 | 0.3280 | 0.2180 | 0.1170 | 0.5190 |
| | | | 5 | *15711.90* | **15822.69** | ***0.1150*** | ***0.1160*** | ***0.2570*** | ***0.1980*** | ***0.1140*** | ***0.4130*** |
| 100 | 50 | MSNPML | 2 | 15787.43 | 15839.57 | 0.1370 | 0.1060 | 0.3470 | 0.1790 | 0.0950 | 0.5410 |
| | | | 3 | 15362.34 | 15434.03 | 0.1460 | 0.1230 | 0.2240 | 0.1360 | 0.0920 | 0.3640 |
| | | | 4 | 15140.14 | 15231.38 | 0.1440 | 0.1310 | 0.1610 | 0.1290 | 0.0873 | 0.2490 |
| | | | 5 | *14995.08* | **15105.87** | ***0.1430*** | ***0.1350*** | ***0.1180*** | ***0.1200*** | ***0.0879*** | ***0.1950*** |
| | | NPML | 2 | 16733.80 | 16785.93 | 0.0813 | 0.0822 | 0.6600 | 0.3430 | 0.0975 | 0.9770 |
| | | | 3 | 16148.48 | 16220.17 | 0.0813 | 0.0823 | 0.4660 | 0.2090 | 0.1020 | 0.8320 |
| | | | 4 | 15752.54 | 15843.78 | 0.0812 | 0.0820 | 0.3420 | 0.1870 | 0.0902 | 0.5390 |
| | | | 5 | *15504.71* | **15615.50** | ***0.0811*** | ***0.0815*** | ***0.2670*** | ***0.1600*** | ***0.0874*** | ***0.4290*** |
| 100 | 100 | MSNPML | 2 | 31750.08 | 31807.76 | 0.1370 | 0.1060 | 0.3660 | 0.1850 | 0.0939 | 0.5550 |
| | | | 3 | 30896.40 | 30975.71 | 0.1430 | 0.1210 | 0.2420 | 0.1410 | 0.0886 | 0.3850 |
| | | | 4 | 30451.67 | 30552.61 | 0.1400 | 0.1330 | 0.1820 | 0.1340 | 0.0871 | 0.2670 |
| | | | 5 | *30133.09* | **30255.67** | ***0.1410*** | ***0.1370*** | ***0.1390*** | ***0.1250*** | ***0.0882*** | ***0.2160*** |
| | | NPML | 2 | 33608.05 | 33665.73 | 0.0807 | 0.0830 | 0.6680 | 0.3510 | 0.0978 | 0.9770 |
| | | | 3 | 32497.87 | 32577.18 | 0.0806 | 0.0830 | 0.4810 | 0.2280 | 0.0967 | 0.8120 |
| | | | 4 | 31704.28 | 31805.22 | 0.0805 | 0.0830 | 0.3590 | 0.1970 | 0.0870 | 0.5430 |
| | | | 5 | *31216.22* | **31338.79** | ***0.0805*** | ***0.0826*** | ***0.2860*** | ***0.1680*** | ***0.0863*** | ***0.4450*** |

Table 3: Simulation results for S3. Italics indicate models with the lowest AIC values, and bold texts indicate models with the lowest BIC values. Columns $\beta_1$, $\beta_2$, $\sigma^2$, $V_{11}$, $V_{12}$, and $V_{13}$ present MAD values of the respective parameter elements, for each model.

| $n$ | $n_j = N$ | Model | $g$ | AIC | BIC | $\beta_1$ | $\beta_2$ | $\sigma^2$ | $V_{11}$ | $V_{12}$ | $V_{22}$ |
|---|---|---|---|---|---|---|---|---|---|---|---|
| 50 | 50 | MSNPML | 2 | 7567.01 | 7613.60 | 0.0749 | 0.0802 | 0.1640 | 0.1520 | 0.1150 | 0.3110 |
|  |  |  | 3 | 7490.64 | 7554.70 | 0.0782 | 0.0883 | 0.1050 | 0.1480 | 0.1130 | 0.2150 |
|  |  |  | 4 | 7448.46 | 7530.00 | 0.0837 | 0.0905 | 0.0686 | 0.1420 | 0.1140 | 0.1810 |
|  |  |  | 5 | *7428.30* | **7527.31** | **0.0841** | **0.0916** | **0.0476** | **0.1400** | **0.1120** | **0.1670** |
|  |  | NPML | 2 | 7850.56 | 7897.15 | 0.1330 | 0.1390 | 0.4200 | 0.3530 | 0.1060 | 0.3520 |
|  |  |  | 3 | 7694.62 | 7758.68 | 0.1330 | 0.1380 | 0.2690 | 0.2230 | 0.1100 | 0.3550 |
|  |  |  | 4 | 7591.19 | 7672.73 | 0.1330 | 0.1380 | 0.1720 | 0.1710 | 0.1150 | 0.3040 |
|  |  |  | 5 | *7534.38* | **7633.39** | **0.1330** | **0.1380** | **0.1290** | **0.1600** | **0.1130** | **0.2550** |
| 50 | 100 | MSNPML | 2 | 15159.42 | 15211.55 | 0.0728 | 0.0731 | 0.1720 | 0.1530 | 0.1080 | 0.3170 |
|  |  |  | 3 | 15010.75 | 15082.44 | 0.0792 | 0.0793 | 0.1190 | 0.1500 | 0.1080 | 0.2220 |
|  |  |  | 4 | 14913.35 | 15004.59 | 0.0786 | 0.0852 | 0.0833 | 0.1440 | 0.1080 | 0.1880 |
|  |  |  | 5 | *14861.26* | **14972.05** | **0.0772** | **0.0838** | **0.0625** | **0.1410** | **0.1070** | **0.1680** |
|  |  | NPML | 2 | 15720.15 | 15772.28 | 0.1400 | 0.1360 | 0.4240 | 0.3590 | 0.1020 | 0.3530 |
|  |  |  | 3 | 15422.99 | 15494.68 | 0.1400 | 0.1360 | 0.2790 | 0.2290 | 0.1050 | 0.3600 |
|  |  |  | 4 | 15194.62 | 15285.86 | 0.1400 | 0.1360 | 0.1840 | 0.1780 | 0.1090 | 0.3150 |
|  |  |  | 5 | *15084.59* | **15195.38** | **0.1400** | **0.1350** | **0.1430** | **0.1640** | **0.1090** | **0.2690** |
| 100 | 50 | MSNPML | 2 | 15032.99 | 15085.13 | 0.0521 | 0.0550 | 0.1730 | 0.1170 | 0.0759 | 0.3160 |
|  |  |  | 3 | 14858.75 | 14930.44 | 0.0576 | 0.0596 | 0.1160 | 0.1140 | 0.0751 | 0.2030 |
|  |  |  | 4 | 14757.16 | 14848.40 | 0.0592 | 0.0630 | 0.0762 | 0.1070 | 0.0755 | 0.1530 |
|  |  |  | 5 | *14697.25* | **14808.04** | **0.0578** | **0.0635** | **0.0508** | **0.1040** | **0.0742** | **0.1280** |
|  |  | NPML | 2 | 15644.37 | 15696.51 | 0.0953 | 0.0963 | 0.4310 | 0.3410 | 0.0719 | 0.3400 |
|  |  |  | 3 | 15300.96 | 15372.65 | 0.0953 | 0.0960 | 0.2840 | 0.1980 | 0.0740 | 0.3520 |
|  |  |  | 4 | 15080.41 | 15171.65 | 0.0953 | 0.0954 | 0.1760 | 0.1300 | 0.0764 | 0.3150 |
|  |  |  | 5 | *14959.83* | **15070.62** | **0.0954** | **0.0952** | **0.1390** | **0.1210** | **0.0765** | **0.2530** |
| 100 | 100 | MSNPML | 2 | 30113.09 | 30170.78 | 0.0500 | 0.0514 | 0.1800 | 0.1190 | 0.0761 | 0.3200 |
|  |  |  | 3 | 29776.37 | 29855.68 | 0.0526 | 0.0571 | 0.1270 | 0.1160 | 0.0762 | 0.2090 |
|  |  |  | 4 | 29547.89 | 29648.83 | 0.0530 | 0.0576 | 0.0891 | 0.1060 | 0.0762 | 0.1600 |
|  |  |  | 5 | *29419.89* | **29542.47** | **0.0525** | **0.0583** | **0.0638** | **0.1040** | **0.0741** | **0.1280** |
|  |  | NPML | 2 | 31309.12 | 31366.80 | 0.0925 | 0.0943 | 0.4350 | 0.3410 | 0.0685 | 0.3380 |
|  |  |  | 3 | 30667.17 | 30746.49 | 0.0925 | 0.0938 | 0.2950 | 0.2010 | 0.0722 | 0.3540 |
|  |  |  | 4 | 30160.96 | 30261.91 | 0.0925 | 0.0936 | 0.1910 | 0.1350 | 0.0751 | 0.3180 |
|  |  |  | 5 | *29938.50* | **30061.08** | **0.0925** | **0.0935** | **0.1520** | **0.1240** | **0.0753** | **0.2640** |

Table 4: Simulation results for S4. Italics indicate models with the lowest AIC values, and bold texts indicate models with the lowest BIC values. Columns $\beta_1$, $\beta_2$, $\sigma^2$, $V_{11}$, $V_{12}$, and $V_{13}$ present MAD values of the respective parameter elements, for each model.

| $n$ | $n_j = N$ | Model | $g$ | AIC | BIC | $\beta_1$ | $\beta_2$ | $\sigma^2$ | $V_{11}$ | $V_{12}$ | $V_{22}$ |
|---|---|---|---|---|---|---|---|---|---|---|---|
| 50 | 50 | MSNPML | 2 | 7566.24 | 7612.83 | 0.0952 | 0.0728 | 0.1430 | 0.1410 | 0.0993 | 0.2420 |
| | | | 3 | 7488.84 | 7552.90 | 0.0986 | 0.0811 | 0.0928 | 0.1360 | 0.0995 | 0.1900 |
| | | | 4 | 7446.49 | 7528.03 | 0.1010 | 0.0861 | 0.0623 | 0.1330 | 0.0982 | 0.1600 |
| | | | 5 | *7426.45* | **7525.46** | **0.1050** | **0.0919** | **0.0453** | **0.1340** | **0.0982** | **0.1480** |
| | | NPML | 2 | 7852.29 | 7898.89 | 0.1290 | 0.1360 | 0.3150 | 0.2590 | 0.0939 | 0.2890 |
| | | | 3 | 7693.69 | 7757.76 | 0.1290 | 0.1360 | 0.2150 | 0.1850 | 0.0950 | 0.2750 |
| | | | 4 | 7589.85 | 7671.39 | 0.1290 | 0.1360 | 0.1500 | 0.1570 | 0.0959 | 0.2390 |
| | | | 5 | *7534.24* | **7633.25** | **0.1290** | **0.1360** | **0.1150** | **0.1450** | **0.0975** | **0.2120** |
| 50 | 100 | MSNPML | 2 | 15158.83 | 15210.97 | 0.0925 | 0.0680 | 0.1530 | 0.1340 | 0.0893 | 0.2420 |
| | | | 3 | 15009.97 | 15081.66 | 0.0987 | 0.0749 | 0.1040 | 0.1280 | 0.0892 | 0.1950 |
| | | | 4 | 14913.87 | 15005.11 | 0.1060 | 0.0785 | 0.0752 | 0.1260 | 0.0877 | 0.1630 |
| | | | 5 | *14862.73* | **14973.52** | **0.1060** | **0.0822** | **0.0586** | **0.1250** | **0.0876** | **0.1500** |
| | | NPML | 2 | 15718.35 | 15770.49 | 0.1200 | 0.1230 | 0.3190 | 0.2540 | 0.0853 | 0.2820 |
| | | | 3 | 15424.38 | 15496.07 | 0.1200 | 0.1230 | 0.2230 | 0.1780 | 0.0867 | 0.2740 |
| | | | 4 | 15198.18 | 15289.42 | 0.1200 | 0.1230 | 0.1620 | 0.1500 | 0.0882 | 0.2430 |
| | | | 5 | *15089.18* | **15199.97** | **0.1200** | **0.1230** | **0.1260** | **0.1390** | **0.0882** | **0.2170** |
| 100 | 50 | MSNPML | 2 | 15032.63 | 15084.77 | 0.0677 | 0.0472 | 0.1490 | 0.1040 | 0.0675 | 0.2470 |
| | | | 3 | 14860.75 | 14932.44 | 0.0719 | 0.0541 | 0.1000 | 0.0978 | 0.0678 | 0.1900 |
| | | | 4 | 14756.06 | 14847.30 | 0.0743 | 0.0617 | 0.0670 | 0.0935 | 0.0665 | 0.1370 |
| | | | 5 | *14698.48* | **14809.27** | **0.0767** | **0.0629** | **0.0477** | **0.0914** | **0.0663** | **0.1190** |
| | | NPML | 2 | 15632.48 | 15684.62 | 0.0873 | 0.0937 | 0.3190 | 0.2480 | 0.0654 | 0.2800 |
| | | | 3 | 15294.46 | 15366.15 | 0.0873 | 0.0938 | 0.2240 | 0.1610 | 0.0653 | 0.2750 |
| | | | 4 | 15076.65 | 15167.89 | 0.0873 | 0.0937 | 0.1540 | 0.1180 | 0.0678 | 0.2470 |
| | | | 5 | *14959.00* | **15069.79** | **0.0872** | **0.0932** | **0.1210** | **0.1070** | **0.0682** | **0.2120** |
| 100 | 100 | MSNPML | 2 | 30115.80 | 30173.48 | 0.0694 | 0.0466 | 0.1560 | 0.1050 | 0.0649 | 0.2480 |
| | | | 3 | 29778.77 | 29858.08 | 0.0752 | 0.0530 | 0.1100 | 0.0970 | 0.0662 | 0.1960 |
| | | | 4 | 29560.85 | 29661.80 | 0.0772 | 0.0578 | 0.0791 | 0.0940 | 0.0651 | 0.1450 |
| | | | 5 | *29429.82* | **29552.40** | **0.0786** | **0.0637** | **0.0608** | **0.0918** | **0.0655** | **0.1240** |
| | | NPML | 2 | 31304.04 | 31361.73 | 0.0908 | 0.0902 | 0.3210 | 0.2430 | 0.0626 | 0.2750 |
| | | | 3 | 30664.49 | 30743.80 | 0.0909 | 0.0902 | 0.2310 | 0.1610 | 0.0624 | 0.2760 |
| | | | 4 | 30159.70 | 30260.65 | 0.0907 | 0.0898 | 0.1650 | 0.1200 | 0.0644 | 0.2470 |
| | | | 5 | *29947.69* | **30070.26** | **0.0908** | **0.0898** | **0.1330** | **0.1080** | **0.0660** | **0.2180** |

We observe that on most occasions throughout the four scenarios, the MSNPML model estimate of the covariance $V_{12}$ is lower than that of the corresponding NPML model estimate, or is close in value. We note that both models appear to estimate the covariance matrix of the random-effects distribution equally well, with very little differences across all cases.

We shall discuss the estimation of the fixed effects $\boldsymbol{\beta}$ in parts. Firstly, we notice that in S3 and S4, the MAD of the MSNPML model estimates are uniformly smaller than those of the NPML model estimates, often by a significant fraction. However, this is different to the results found in S1, where the MAD of the MSNPML model estimates are very close to those of the NPML model estimates, often slightly greater; and in S2, the MAD of the MSNPML model estimates are uniformly greater that those of the NPML model estimates. We offer the following reasons for these observations.

In Figure 1, we observe that the S3 and S4 samples clearly arise from marginally symmetric distributions, due to the clustered nature of the distributions and the separations between the modes. This is consistent with the results from Agresti et al. (2004). However, in S1 and S2, a small sample can often appear skewed, as is most evident in S2 panel of Figure 1.

Thus, the small sample skewness can be better modeled by the flexibility of the NPML model, since it does not enforce symmetry on the estimated density. The MSNPML model estimates tend to model the outlier random effects, in small samples, thus shifting the fixed-effects parameter elements away from their true value. However, when the random effects exhibit symmetry in small samples, there are less biasing effects on the fixed-effects estimates of the MSNPML model, as is observable in the results of Tables 3 and 4.

Further, we observe that under both models, an increase in $g$ results in decreases in the MAD of the estimates for the random-effects covariance matrix and the error variance. However, there appears to be little effects of increasing $g$ on the MAD of the fixed-effects estimates. This is in agreement with the results reported in Agresti et al. (2004), Butler and Louis (1992), and McCulloch and Neuhaus (2011). We observe that there is an decrease in the MAD of the fixed effects due to increase in sample size, across all four scenarios, which is in line with general asymptotic theory.

The MAD results for simulations S5 and S6 are reported in Tables 5 and 6, respectively. In both scenarios, we notice that NPML uniformly outperforms MSNPML with respect to the estimation of the fixed effects. However, the estimated efficiency for $V_{11}$ of both methods appear to be similar in S5 and S6. Surprisingly, MSNPML is uniformly more efficient at estimating the error variance $\sigma^2$. Furthermore, in both scenarios, MSNPML produces more efficient estimates of $V_{22}$. These results indicate that MSNPML may still be useful for estimating variance components, even when the random-effects distribution is not symmetric.

We report the AIC and BIC selection results for scenarios S1–S6 in Tables 1–6, respectively. We observe that, in general, the AIC and BIC preferred the most complex models (i.e. $g = 5$) for both MSNPML and NPML, across all $n$ and $N$ values, with few exception. That is, in S1, the BIC preferred $g = 4$ for both MSNPML and NPML, when $n = 50$ and $N = 50$, and MSNPML only, when $n = 50$ and $N = 100$; in S5 and S6, the BIC preferred $g = 4$, when $n = 50$ and $N = 50$. Further, across all scenarios, both the AIC and BIC preferred the MSNPML models to the NPML models, across all numbers of components, and with respect to the models that are preferred over all values of $g$. In scenarios S1–S4, the selected MSNPML models uniformly outperform the

Table 5: Simulation results for S5. Italics indicate models with the lowest AIC values, and bold texts indicate models with the lowest BIC values. Columns $\beta_1$, $\beta_2$, $\sigma^2$, $V_{11}$, $V_{12}$, and $V_{13}$ present MAD values of the respective parameter elements, for each model.

| $n$ | $n_j = N$ | Model | $g$ | AIC | BIC | $\beta_1$ | $\beta_2$ | $\sigma^2$ | $V_{11}$ | $V_{12}$ | $V_{22}$ |
|---|---|---|---|---|---|---|---|---|---|---|---|
| 50 | 50 | MSNPML | 2 | 7479.32 | 7525.91 | 0.1100 | 0.1002 | 0.1012 | 0.0711 | 0.0575 | 0.1650 |
| | | | 3 | 7404.58 | 7468.65 | 0.1092 | 0.1110 | 0.0529 | 0.0601 | 0.0553 | 0.0958 |
| | | | 4 | 7376.42 | **7457.96** | **0.1101** | **0.1173** | **0.0337** | **0.0585** | **0.0529** | **0.0740** |
| | | | 5 | *7363.95* | 7462.96 | *0.1131* | *0.1234* | *0.0261* | *0.0583* | *0.0531* | *0.0694* |
| | | NPML | 2 | 7676.80 | 7723.39 | 0.0692 | 0.0782 | 0.2233 | 0.0831 | 0.0566 | 0.3346 |
| | | | 3 | 7550.29 | 7614.35 | 0.0692 | 0.0785 | 0.1435 | 0.0651 | 0.0563 | 0.2259 |
| | | | 4 | 7471.25 | 7552.78 | 0.0692 | 0.0778 | 0.0938 | 0.0575 | 0.0527 | 0.1307 |
| | | | 5 | *7428.58* | **7527.59** | **0.0691** | **0.0782** | **0.0650** | **0.0545** | **0.0512** | **0.0903** |
| 50 | 100 | MSNPML | 2 | 14986.74 | 15038.88 | 0.0903 | 0.0737 | 0.1111 | 0.0599 | 0.0432 | 0.1757 |
| | | | 3 | 14831.14 | 14902.83 | 0.0947 | 0.0836 | 0.0613 | 0.0460 | 0.0402 | 0.1017 |
| | | | 4 | 14767.86 | 14859.10 | 0.0967 | 0.0865 | 0.0382 | 0.0441 | 0.0396 | 0.0689 |
| | | | 5 | *14735.12* | **14845.91** | **0.0965** | **0.0901** | **0.0259** | **0.0427** | **0.0396** | **0.0545** |
| | | NPML | 2 | 15363.23 | 15415.36 | 0.0529 | 0.0563 | 0.2292 | 0.0776 | 0.0413 | 0.3441 |
| | | | 3 | 15130.17 | 15201.86 | 0.0528 | 0.0560 | 0.1552 | 0.0523 | 0.0437 | 0.2524 |
| | | | 4 | 14972.10 | 15063.34 | 0.0527 | 0.0559 | 0.1062 | 0.0433 | 0.0399 | 0.1426 |
| | | | 5 | *14881.98* | **14992.77** | **0.0527** | **0.0557** | **0.0770** | **0.0398** | **0.0379** | **0.0951** |
| 100 | 50 | MSNPML | 2 | 14873.65 | 14925.79 | 0.1091 | 0.0962 | 0.1131 | 0.0731 | 0.0526 | 0.1784 |
| | | | 3 | 14700.39 | 14772.08 | 0.1138 | 0.1046 | 0.0659 | 0.0612 | 0.0504 | 0.1117 |
| | | | 4 | 14622.78 | 14714.02 | 0.1156 | 0.1146 | 0.0433 | 0.0583 | 0.0494 | 0.0822 |
| | | | 5 | *14578.67* | **14689.47** | **0.1126** | **0.1149** | **0.0305** | **0.0572** | **0.0492** | **0.0703** |
| | | NPML | 2 | 15300.11 | 15352.25 | 0.0746 | 0.0738 | 0.2291 | 0.0875 | 0.0533 | 0.3369 |
| | | | 3 | 15032.10 | 15103.79 | 0.0746 | 0.0736 | 0.1547 | 0.0665 | 0.0526 | 0.2406 |
| | | | 4 | 14854.35 | 14945.59 | 0.0745 | 0.0733 | 0.1069 | 0.0575 | 0.0485 | 0.1452 |
| | | | 5 | *14752.56* | **14863.35** | **0.0746** | **0.0733** | **0.0788** | **0.0538** | **0.0472** | **0.1058** |
| 100 | 100 | MSNPML | 2 | 29827.16 | 29884.84 | 0.0895 | 0.0705 | 0.1237 | 0.0612 | 0.0396 | 0.1883 |
| | | | 3 | 29464.43 | 29543.74 | 0.0949 | 0.0827 | 0.0751 | 0.0459 | 0.0358 | 0.1156 |
| | | | 4 | 29298.52 | 29399.46 | 0.0952 | 0.0863 | 0.0514 | 0.0441 | 0.0340 | 0.0783 |
| | | | 5 | *29201.46* | **29324.04** | **0.0980** | **0.0904** | **0.0367** | **0.0419** | **0.0346** | **0.0591** |
| | | NPML | 2 | 30659.76 | 30717.45 | 0.0476 | 0.0522 | 0.2381 | 0.0805 | 0.0379 | 0.3457 |
| | | | 3 | 30151.25 | 30230.57 | 0.0476 | 0.0520 | 0.1672 | 0.0525 | 0.0391 | 0.2617 |
| | | | 4 | 29791.10 | 29892.05 | 0.0474 | 0.0517 | 0.1189 | 0.0432 | 0.0357 | 0.1527 |
| | | | 5 | *29578.60* | **29701.17** | **0.0475** | **0.0518** | **0.0902** | **0.0389** | **0.0343** | **0.1091** |

Table 6: Simulation results for S6. Italics indicate models with the lowest AIC values, and bold texts indicate models with the lowest BIC values. Columns $\beta_1$, $\beta_2$, $\sigma^2$, $V_{11}$, $V_{12}$, and $V_{13}$ present MAD values of the respective parameter elements, for each model.

| $n$ | $n_j = N$ | Model | $g$ | AIC | BIC | $\beta_1$ | $\beta_2$ | $\sigma^2$ | $V_{11}$ | $V_{12}$ | $V_{22}$ |
|---|---|---|---|---|---|---|---|---|---|---|---|
| 50 | 50 | MSNPML | 2 | 7494.42 | 7541.01 | 0.1572 | 0.1269 | 0.1137 | 0.1190 | 0.0542 | 0.1897 |
|  |  |  | 3 | 7410.62 | 7474.68 | 0.1516 | 0.1283 | 0.0622 | 0.1084 | 0.0506 | 0.1275 |
|  |  |  | 4 | 7375.15 | **7456.68** | **0.1556** | **0.1291** | **0.0395** | **0.1085** | **0.0526** | **0.1078** |
|  |  |  | 5 | *7358.18* | 7457.19 | *0.1529* | *0.1245* | *0.0295* | *0.1081* | *0.0490* | *0.1002* |
|  |  | NPML | 2 | 7695.82 | 7742.41 | 0.0676 | 0.0724 | 0.2396 | 0.1557 | 0.0521 | 0.3264 |
|  |  |  | 3 | 7548.14 | 7612.20 | 0.0674 | 0.0714 | 0.1494 | 0.1150 | 0.0538 | 0.2466 |
|  |  |  | 4 | 7467.09 | 7548.63 | 0.0674 | 0.0711 | 0.0990 | 0.1032 | 0.0473 | 0.1707 |
|  |  |  | 5 | *7419.41* | **7518.42** | ***0.0674*** | ***0.0710*** | ***0.0683*** | ***0.0990*** | ***0.0444*** | ***0.1367*** |
| 50 | 100 | MSNPML | 2 | 15052.43 | 15104.57 | 0.1213 | 0.1001 | 0.1321 | 0.0964 | 0.0393 | 0.1996 |
|  |  |  | 3 | 14878.34 | 14950.03 | 0.1185 | 0.0894 | 0.0792 | 0.0785 | 0.0356 | 0.1225 |
|  |  |  | 4 | 14793.62 | 14884.86 | 0.1172 | 0.0850 | 0.0519 | 0.0793 | 0.0359 | 0.0944 |
|  |  |  | 5 | *14744.08* | **14854.87** | ***0.1139*** | ***0.0817*** | ***0.0350*** | ***0.0740*** | ***0.0342*** | ***0.0787*** |
|  |  | NPML | 2 | 15446.48 | 15498.62 | 0.0501 | 0.0526 | 0.2565 | 0.1505 | 0.0386 | 0.3409 |
|  |  |  | 3 | 15167.99 | 15239.68 | 0.0501 | 0.0523 | 0.1714 | 0.0984 | 0.0434 | 0.2799 |
|  |  |  | 4 | 15002.02 | 15093.27 | 0.0499 | 0.0524 | 0.1203 | 0.0875 | 0.0336 | 0.1811 |
|  |  |  | 5 | *14896.33* | **15007.12** | ***0.0498*** | ***0.0520*** | ***0.0879*** | ***0.0778*** | ***0.0328*** | ***0.1432*** |
| 100 | 50 | MSNPML | 2 | 15031.27 | 15083.77 | 0.1680 | 0.1354 | 0.1297 | 0.1229 | 0.0559 | 0.1989 |
|  |  |  | 3 | 14838.98 | 14911.17 | 0.1631 | 0.1292 | 0.0787 | 0.1116 | 0.0520 | 0.1306 |
|  |  |  | 4 | 14747.16 | 14839.05 | 0.1726 | 0.1319 | 0.0531 | 0.1212 | 0.0522 | 0.1115 |
|  |  |  | 5 | *14691.69* | **14803.26** | ***0.1734*** | ***0.1310*** | ***0.0374*** | ***0.1127*** | ***0.0508*** | ***0.1023*** |
|  |  | NPML | 2 | 15460.90 | 15513.40 | 0.0683 | 0.0706 | 0.2475 | 0.1553 | 0.0503 | 0.3277 |
|  |  |  | 3 | 15139.51 | 15211.63 | 0.0682 | 0.0703 | 0.1642 | 0.1174 | 0.0520 | 0.2540 |
|  |  |  | 4 | 14971.17 | 15063.05 | 0.0682 | 0.0701 | 0.1150 | 0.1061 | 0.0454 | 0.1799 |
|  |  |  | 5 | *14855.60* | **14967.18** | ***0.0682*** | ***0.0702*** | ***0.0840*** | ***0.0989*** | ***0.0440*** | ***0.1451*** |
| 100 | 100 | MSNPML | 2 | 30028.92 | 30086.72 | 0.1279 | 0.1012 | 0.1469 | 0.0966 | 0.0367 | 0.2084 |
|  |  |  | 3 | 29636.45 | 29715.93 | 0.1222 | 0.1024 | 0.0954 | 0.0797 | 0.0351 | 0.1338 |
|  |  |  | 4 | 29443.06 | 29544.20 | 0.1281 | 0.0948 | 0.0690 | 0.0768 | 0.0343 | 0.1050 |
|  |  |  | 5 | *29319.17* | **29441.99** | ***0.1282*** | ***0.0938*** | ***0.0515*** | ***0.0765*** | ***0.0332*** | ***0.0886*** |
|  |  | NPML | 2 | 30850.12 | 30907.91 | 0.0472 | 0.0499 | 0.2608 | 0.1539 | 0.0351 | 0.3432 |
|  |  |  | 3 | 30295.06 | 30374.53 | 0.0471 | 0.0498 | 0.1838 | 0.1054 | 0.0408 | 0.2823 |
|  |  |  | 4 | 29923.90 | 30025.05 | 0.0471 | 0.0494 | 0.1350 | 0.0940 | 0.0318 | 0.1943 |
|  |  |  | 5 | *29688.32* | **29811.15** | ***0.0471*** | ***0.0492*** | ***0.1039*** | ***0.0824*** | ***0.0307*** | ***0.1582*** |

respective NPML models, with respect to the MAD of the covariance matrix and error variance. However, in S1, the preferred MSNPML and NPML perform equally with respect to the fixed effects, and in S2, was uniformly more efficient. In S5, the MADs for $V_{11}$ and $V_{12}$ in the NPML models tended to be smaller than those of the respective MSNPML models, but the MAD for $\sigma^2$ and $V_{22}$ were uniformly larger for each $n$ and $N$. In S6, the MADs for $V_{11}$ and $V_{12}$ in both models appear to be similar; however the MAD for $\sigma^2$ and $V_{22}$ were uniformly larger for NPML than MSNPML models, again.

The AIC and BIC results from Tables 1–6 are not surprising, considering that the error variance is smaller for MSNPML models compared to the respective NPML models, across all scenarios in Tables 1–6. This indicates that MSNPML tends to fit the data better overall, even when it is less efficient for estimating the fixed effects.

Finally, we observe that across scenarios S1–S4, there is little difference between the MAD of covariance matrix estimates when using the MSNPML model with $g = 2$ and that of the NPML model with $g = 4$. Thus, we observe that when the random effects are symmetric, the MSNPML model and the NPML model with the same number of point masses, perform similarly. Furthermore, upon inspecting the AIC and BIC values from Tables 1–6, we observe that the corresponding values for $g = 2$ MSNPML models, and $g = 4$ NPML models are highly comparable. Given that the NPML model requires twice the number of random-effects parameter elements for the same number of point masses, this implies the MSNPML model confers an economy benefit in computational requirement and parsimony of specification in cases where the random-effects densities are marginally symmetric. Additionally, the evidence from Tables 5 and 6 suggest that MSNPML may still be useful even when there are deviations away from symmetry.

## 5 Example Applications

In this section, we demonstrate the use of the MSNPML LMM on a pair of data sets from the *nlme R* package (Pinheiro et al., 2014). The data sets that we use as examples are the Female subset of the *Orthodont* data set (Pinheiro and Bates, 2000, App. A.17) and the *Oxboy* data set (Pinheiro and Bates, 2000, App. A.19). The analyses are as follows.

### 5.1 Orthodont

The Female subset of the *Orthodont* data set (which we shall henceforth refer to as the *Orthodont* data set) consists of the measurements of distance between the pituitary and the pterygomaxillary fissure (in millimeters) for $n = 11$ girls, at $n_j = N = 4$ time points: ages 8, 10, 12, and 14 years old. The use of the measurement as a proxy for overall growth is due to the ease of identifiability of the features in x-ray images. The data were originally studied in Potthoff and Roy (1964), and reanalyzed in Pinheiro and Bates (2000, Sec. 1.4) as an example.

In our analysis, we study the data using the same model as applied in Pinheiro and Bates (2000, Sec. 1.4). In reference to (1), we set $y_{jk}$ to the distance, and $\boldsymbol{x}_{jk} = (1, \text{age}_k)^T$ and $\boldsymbol{z}_{jk} = 1$, for each girl $j = 1, ..., 11$ and time points $k = 1, ..., 4$, where $\text{age}_k = 6 + 2k$; this is an analysis of covariance model with a random intercept $\boldsymbol{b}_j = b_j$ for each girl $j$.

Using the AIC/BIC rule, we selected an MSNPML LMM with $2g = 4$ point masses, where the AIC/BIC value was 143.79/146.17; the AIC/BIC values for the $2g = 2$ and $2g = 6$ cases were 176.47/178.06 and 147.68/150.86, respectively. Over the range $g = 2, ..., 6$, the NPML LMMs yielded AIC/BIC values of 175.69/177.28, 163.63/166.02, 146.77/149.95, 150.77/154.75, and 154.77/159.54, respectively. Thus the MSNPML can be seen as being a better fit for this data.

Table 7: Parameter estimates for the components of $\boldsymbol{\theta}$ in the $2g = 4$ MSNPML LMM for the *Orthodont* data set.

| Parameter | $\boldsymbol{\beta} = \begin{bmatrix} \beta_1 \\ \beta_2 \end{bmatrix}$ | $\sigma^2$ | $\pi_1$ | $\pi_2$ | $\lambda_1$ | $\lambda_2$ |
|---|---|---|---|---|---|---|
| Estimate | $\begin{bmatrix} 16.87 \\ 0.4795 \end{bmatrix}$ | 0.6182 | 0.7274 | 0.2726 | $-0.9782$ | $-3.534$ |
| Standard Error | $\begin{bmatrix} (0.5961) \\ (0.05301) \end{bmatrix}$ | (0.1323) | (0.1343) | — | (0.2308) | (0.1427) |

Under the $2g = 4$ MSNPML LMM, the parameter estimates are listed in Table 7. In order to calculate the standard errors, we utilized a numerical approximation of the information matrix $\boldsymbol{I}(\boldsymbol{\theta})$, evaluated at the parameter estimate $\hat{\boldsymbol{\theta}}_n$; the numerical differentiation was conducted using the *numDeriv* R package (Gilbert and Varadhan, 2012).

The random-effects variance $\text{var}(B) = V = \pi_1 \lambda_1^2 + \pi_2 \lambda_2^2$ can be estimated as $\hat{V} = 4.101$, with a standard error of 1.626. Here, we obtain the standard error via the Delta method (e.g. see Theorem 1.3 of Boos and Stefanski (2013)), which states that $\hat{V}$ asymptotically has an approximately normal distribution with mean $\text{var}(B)$ and variance $n^{-1} \nabla V(\boldsymbol{\theta}_0) \boldsymbol{I}^{-1}(\boldsymbol{\theta}_0) [\nabla V(\boldsymbol{\theta}_0)]^T$, under the assumption that $\hat{\boldsymbol{\theta}}_n$ is a consistent estimator; since $\boldsymbol{\theta}_0$ is unknown, the variance can be estimated by evaluation at $\hat{\boldsymbol{\theta}}_n$. As expected, the 95% confidence interval of $V$, $(0.9140, 7.288)$, indicates that there is heterogeneity in the individual growth patterns of the girls in the data.

Here, we utilized confidence intervals as an approximate inferential heuristic for assessing the necessity of the random-effects model. As noted by a reviewer, it is well known that the test of $V = 0$ versus $V \neq 0$ is difficult and cannot be performed using general asymptotic theory (cf. Stram and Lee (1994)). As suggested by a reviewer, we also computed a 95% bootstrapped percentile interval (cf. Efron and Tibshirani (1993, Ch. 13)) using a parametric bootstrap via in the style of McLachlan (1987). The resulting interval, $(1.266, 7.399)$, compares favorably with the asymptotic interval that we have reported.

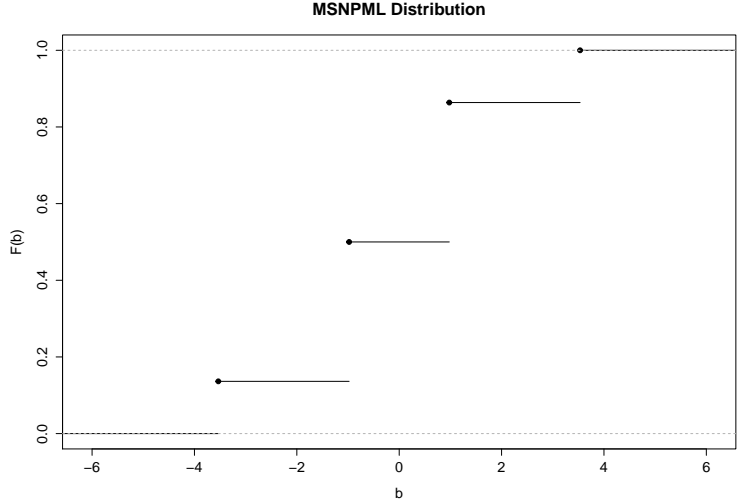

Figure 2: Estimated distribution function with $2g = 4$ point masses, for the *Orthodont* data set.

To visually portray the results, we plot the MSNPML distribution function in Figure 2. A plot of the empirical distribution function of the estimated *a posteriori* expectations of the random effects $\hat{b}_j$ (as calculated using equation (17)) appears in Figure 3. By comparing Figures 2 and 3, we note that although the MSNPML distribution can only take four distinct values, the possible values of $\hat{b}_j$ can potentially be any convex combination of the point masses. Furthermore, unlike the random-effects distribution, $\hat{b}_j$ need not be symmetrically distributed.

Suppose that we wish to predict the unknown response $Y_j^*$ of individual $j$, where we only know the covariates $\boldsymbol{x}_j^*$ and $\boldsymbol{z}_j^*$. Given estimates $\hat{\boldsymbol{b}}_j$ and $\hat{\boldsymbol{\beta}}_n$, we can write the estimated expectation of $Y_j^*$ as

$$\hat{\mathbb{E}}\left(Y_j^* | \boldsymbol{B} = \hat{\boldsymbol{b}}_j\right) = \hat{\boldsymbol{\beta}}_n^T \boldsymbol{x}_j^* + \hat{\boldsymbol{b}}_j^T \boldsymbol{z}_j^*. \tag{19}$$

We shall refer to (19) as the individual estimated expectation of $j$. The individual estimated expectation line for each of the 11 girls is plotted in Figure 4, for $\text{age}_j \in (8, 14)$. As expected, we observe that the estimated expectation lines are all parallel (due to there being no random effects on age), and are clustered around four point masses, as was visually evident in Figure 3. Regardless, the clustered estimated lines are fairly well fitted to the data of each individual, as only the green points are missed entirely.

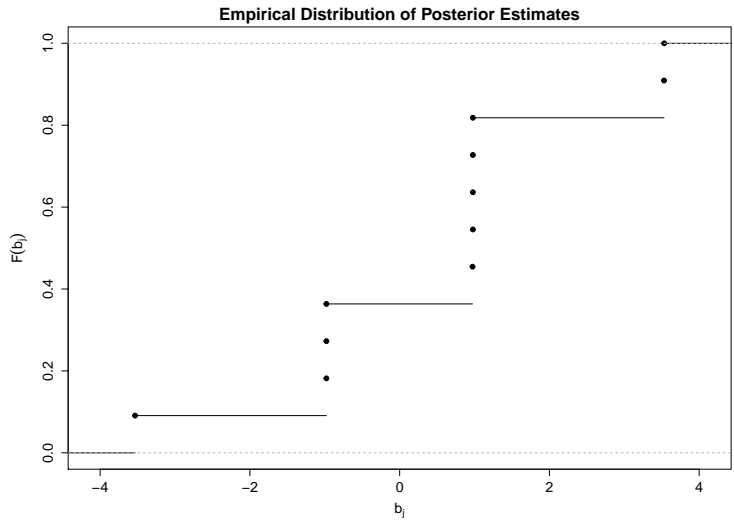

Figure 3: Empirical distribution function of the *a posteriori* expectations of the random effects $\hat{b}_j$, estimated from the MSNPML model, for the *Orthodont* data set.

To conclude the analysis of the *Orthodont* data set, we present the parameter estimates from the analysis of Pinheiro and Bates (2000, Sec. 1.4) in Table 8; here, the authors utilize a conventional LMM with a normal random-effects model (i.e. $f_B(b) = \phi(b; 0, \sigma_B^2)$). We notice that the estimates from Table 8 are very similar to that of the MSNPML LMM. Furthermore, we observe that the standard errors of the intercept estimate is slightly larger than the ones from Table 7. This is unsurprising, as the standard errors obtained from Table 8 are finite sample corrected and are therefore more conservative than our asymptotic standard errors. Finally, we note that the AIC and BIC for the normal LMM are 149.22 and 156.17, respectively. This indicates that the MSNPML LMM achieved a better fit than the usual normal model in this data set.

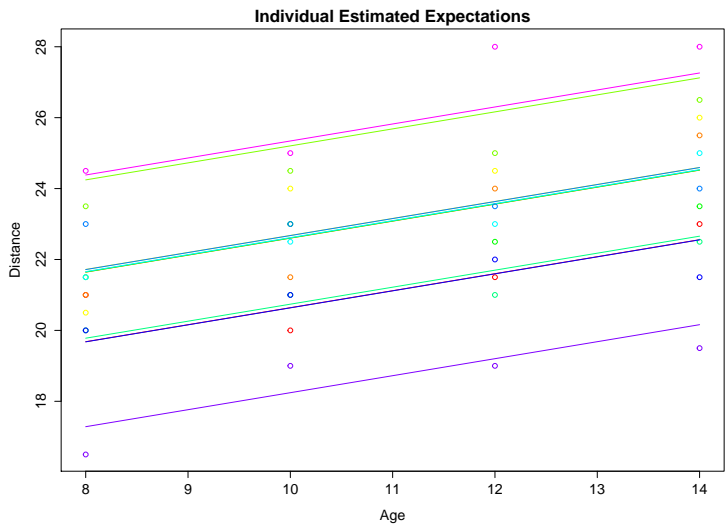

Figure 4: Individual estimated expectation lines for each of the 11 girls from the the *Orthodont* data set. Each color indicates a different individual.

Table 8: Parameter estimates of the LMM model for the *Orthodont* data set, from Pinheiro and Bates (2000, Sec. 1.4). Here, $\sigma_B^2$ is the variance of the normal random-effects model. Dashes indicate where no information was provided.

| Parameter | $\boldsymbol{\beta} = \begin{bmatrix} \beta_1 \\ \beta_2 \end{bmatrix}$ | $\sigma^2$ | $\sigma_B^2$ |
|---|---|---|---|
| Estimate | $\begin{bmatrix} 17.373 \\ 0.480 \end{bmatrix}$ | 0.6084 | 4.278 |
| Standard Error | $\begin{bmatrix} (0.85874) \\ (0.05259) \end{bmatrix}$ | — | — |

## 5.2 Oxboy

The *Oxboy* data set contains the height measurements (in centimeters) of $n = 26$ boys from Oxford, England. The boys are each measured at $n_j = N = 9$ time points which were converted to a standardized age (an age measurement approximately between $-1$ and 1).

For each boy $j = 1, ..., 26$ and each time of measurement $k = 1, ..., 9$, we set $y_{jk}$ to be the measured height and $\boldsymbol{x}_{jk} = \boldsymbol{z}_{jk} = (1, \text{age}_{jk})^T$, where $\text{age}_{jk}$ is the standardized age of boy $j$ at time of measurement $k$. For our analysis, we fit an MSNPML LMM with random effects on both the intercept and the slope of the regression between height and age.

Using the AIC/BIC rule, we selected an MSNPML LMM with $2g = 16$ point masses, where the AIC/BIC value was 843.61/876.32; the AIC/BIC values for the $2g = 14$ and $2g = 18$ cases were 883.95/912.88 and 874.34/910.82, respectively. Over the range $g = 14, ..., 18$, the NPML LMMs yielded AIC/BIC values of 1104.96/1156.55, 978.89/1034.24, 907.19/966.32, 914.28/977.19, and 1122.63/1189.31, respectively. Thus, like in the *Orthodont* data, the MSNPML can be seen as being a better fit for this data as well.

Under the $2g = 16$ MSNPML LMM, the parameter estimates are as appear in Table 9. Further, using Equation (16) and the notation of (18), we obtain the estimates for the distinct elements of the random-effects covariance matrix to give $\hat{V}_{11} = 62.64$ (17.20), $\hat{V}_{12} = 8.253$ (3.053), and $\hat{V}_{22} = 2.550$ (0.5558); again, the standard errors reported in the parentheses are obtained via the Delta method. Considering the sizes of the standard errors as compared to the estimates, we can conclude that there is likely heterogeneity in the growth patterns of the boys in the data.

Here, we note again that our conclusion should be taken to be a heuristic assessment, as opposed to formal tests of hypotheses. For completeness, the 95% bootstrapped percentile intervals for $V_{11}$, $V_{12}$, and $V_{22}$ are $(33.50, 99.65)$, $(1.970, 14.30)$, and $(1.186, 3.139)$, respectively. These intervals are very similar to those that would be obtained via the standard errors from the Delta method.

In Figure 5, we plot the point masses and the estimates $\hat{\boldsymbol{b}}_j$, in order to visualize the *a priori* random-effects density and the density of the estimated

Table 9: Parameter estimates for the components of $\boldsymbol{\theta}$ in the $2g = 16$ MSNPML LMM for the *Oxboy* data set.

| Parameter | $\boldsymbol{\beta} = \begin{bmatrix} \beta_1 \\ \beta_2 \end{bmatrix}$ | $\sigma^2$ |
|---|---|---|
| Estimate | $\begin{bmatrix} 148.9 \\ 6.885 \end{bmatrix}$ | 0.9431 |
| Standard Error | $\begin{bmatrix} (0.08751) \\ (0.13100) \end{bmatrix}$ | (0.08785) |

| Parameter | $\pi_1$ | $\pi_2$ | $\pi_3$ | $\pi_4$ | $\pi_5$ | $\pi_6$ | $\pi_7$ | $\pi_8$ |
|---|---|---|---|---|---|---|---|---|
| Estimate | 0.1538 | 0.1547 | 0.07692 | 0.1538 | 0.1154 | 0.1516 | 0.07695 | 0.11683 |
| Standard Error | (0.07076) | (0.07320) | (0.05225) | (0.05227) | (0.06265) | (0.07076) | (0.06436) | — |

| Parameter | $\boldsymbol{\lambda}_1 = \begin{bmatrix} \lambda_{11} \\ \lambda_{12} \end{bmatrix}$ | $\boldsymbol{\lambda}_2 = \begin{bmatrix} \lambda_{21} \\ \lambda_{22} \end{bmatrix}$ | $\boldsymbol{\lambda}_3 = \begin{bmatrix} \lambda_{31} \\ \lambda_{32} \end{bmatrix}$ | $\boldsymbol{\lambda}_4 = \begin{bmatrix} \lambda_{41} \\ \lambda_{42} \end{bmatrix}$ | $\boldsymbol{\lambda}_5 = \begin{bmatrix} \lambda_{51} \\ \lambda_{52} \end{bmatrix}$ | $\boldsymbol{\lambda}_6 = \begin{bmatrix} \lambda_{61} \\ \lambda_{62} \end{bmatrix}$ | $\boldsymbol{\lambda}_7 = \begin{bmatrix} \lambda_{71} \\ \lambda_{72} \end{bmatrix}$ | $\boldsymbol{\lambda}_8 = \begin{bmatrix} \lambda_{81} \\ \lambda_{82} \end{bmatrix}$ |
|---|---|---|---|---|---|---|---|---|
| Estimate | $\begin{bmatrix} -6.708 \\ -1.096 \end{bmatrix}$ | $\begin{bmatrix} 1.054 \\ 0.5812 \end{bmatrix}$ | $\begin{bmatrix} -6.336 \\ 1.919 \end{bmatrix}$ | $\begin{bmatrix} -10.47 \\ -1.705 \end{bmatrix}$ | $\begin{bmatrix} 16.84 \\ 2.606 \end{bmatrix}$ | $\begin{bmatrix} -2.409 \\ 0.9625 \end{bmatrix}$ | $\begin{bmatrix} -4.433 \\ 0.01794 \end{bmatrix}$ | $\begin{bmatrix} 2.409 \\ -0.9625 \end{bmatrix}$ |
| Standard Error | $\begin{bmatrix} (0.1678) \\ (0.2592) \end{bmatrix}$ | $\begin{bmatrix} (0.1821) \\ (0.2690) \end{bmatrix}$ | $\begin{bmatrix} (0.2290) \\ (0.3541) \end{bmatrix}$ | $\begin{bmatrix} (0.2295) \\ (0.3545) \end{bmatrix}$ | $\begin{bmatrix} (0.1893) \\ (0.2924) \end{bmatrix}$ | $\begin{bmatrix} (0.1678) \\ (0.2585) \end{bmatrix}$ | $\begin{bmatrix} (0.2351) \\ (0.3394) \end{bmatrix}$ | $\begin{bmatrix} (0.1841) \\ (0.2991) \end{bmatrix}$ |

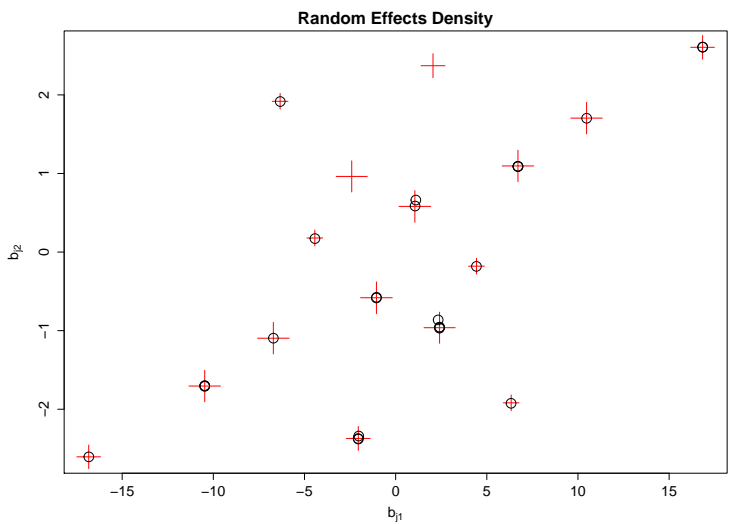

Figure 5: The estimated random-effect density for the *Oxboy* data set is plotted as plus signs, where the size is relative to the weight of that point mass (see Table 9). The circles are each of individual estimated *a posteriori* expectations $\hat{\boldsymbol{b}}_j$.

*a posteriori* expectations. We note from the figure that the positive correlation observed in the estimated covariance matrix is visually evident. Further, we once again see that the estimated *a posteriori* expectations are clustered around the point masses, with few deviating any significant distance from the nearest point mass. Unlike in the *Orthodont* data, however, we observe that some of the point masses have no nearby *a posteriori* expectation estimates. This is likely due to the large proportions of estimates around their respective mirrored masses.

Lastly, we plot the individual estimated expectation lines for each of the 26 boys in Figure 6. Unlike in Figure 4, the estimated lines are no longer parallel since there is are now random effects on the slope as well as the intercept. We see that the lines are fairly well fitted to each individual's data, especially in the centre of the distribution. However, there is some lack of fit in the peripheries. For example, the green points are missed all together at the bottom of the graph.

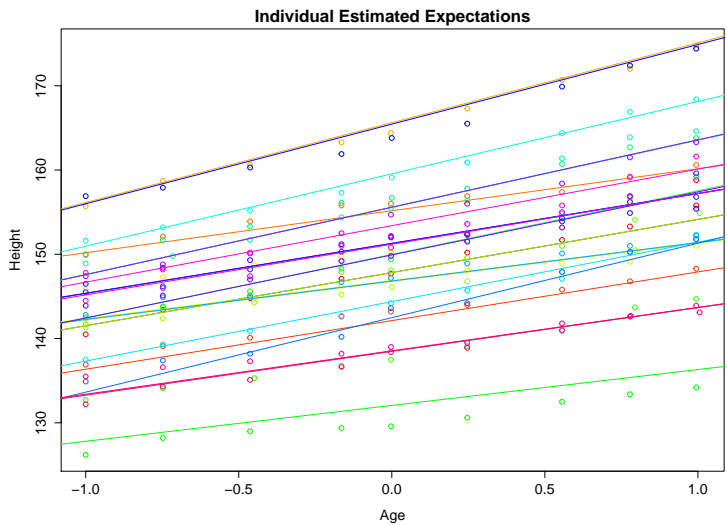

Figure 6: Individual estimated expectation lines for each of the 26 boys from the the *Oxboy* data set. Each color indicates a different individual.

## 6 Conclusions

In this article, we have introduced the MSNPML LMM for modeling random effects under the assumption of marginal symmetry. Under such conditions, the MSNPML model utilizes half the number of parameters to model the same number of point masses as the NPML model.

An ECM algorithm is presented for the ML estimation of the parameters of the MSNPML LMM. The algorithm is shown to exhibit monotonicity in the sequence of iterate likelihoods, as well as convergent to a stationary point of the log-likelihood function. Furthermore, the ML estimator is shown to be consistent and asymptotically normal in balanced experiments and when covariates are random. Conditions are discussed regarding the necessary conditions in the case of fixed covariates. The estimation of interesting quantities such as the random-effects covariance matrix and the individual *a posteriori* expectations is discussed, as well as the selection of the number of point masses via the AIC or BIC rules.

We then demonstrated that the MSNPML model was more efficient than the NPML model, via a set of four numerical simulations. In these simulations, it was found that the MSNPML model was better able to estimate accurately the random-effects marginal variances and error variance in all settings. The estimation of random-effects covariance matrix appear to be similar in both models across all cases. In the cases where the random-effects distributions are obviously marginally symmetric in small samples, the MSNPML model was more efficient in estimating the fixed effects. However, in the cases where there is possible skewness in small samples, the NPML model can outperform the MSNPML model.

A pair of examples from Pinheiro and Bates (2000) was used to demonstrate the manner in which an MSNPML model could be used to conduct statistical inference. In the *Orthodont* data set, it was found that the inferences produced via the MSNPML model corresponded closely to those obtained via the normal LMM model (cf. Pinheiro and Bates (2000, Sec. 1.4)). In the *Oxboy* data set, we demonstrated how the model could be used to draw inferences regarding the heterogeneity in the data set. We also demonstrated a variety of visualization tools for graphically interpreting MSNPML model estimates.

In the future, it would be interesting to extend the MSNPML model to LMM cases where the error variances are not normal. For instance, we could use the mixture regression results of Nguyen and McLachlan (2016) or Galimberti and Soffritti (2014) to estimate MSNPML LMMs with Laplace or $t$ distributed errors, respectively. It is also possible to allow for skewed error distributions via the mixture results of Lee and McLachlan (2016).

Further, it is possible to extend the model to the so called generalized LMM context. We note that this was explored in part by Agresti et al. (2004) and Aitkin (1999) in the case of NPML models. However, in neither articles are algorithms presented that exhibit the monotonicity properties of proper EM-type algorithms. It is possible to modify the estimation techniques for mixtures of experts from Nguyen and McLachlan (2016) for use in estimating NPML and MSNPML multinomial regressions. Additionally, we may also explore algorithms for applying NPML and MSNPML models in other generalized settings such as in beta and gamma regressions.

# References

Agresti, A., Caffo, B., Ohman-Strickland, P., 2004. Examples in which misspecification of a random effects distribution reduces efficiency, and possible remedies. Computational Statistics and Data Analysis 47, 639–653.

Aitkin, M., 1999. A general maximum likelihood analysis of variance components in generalized linear models. Biometrics 55, 117–128.

Akaike, H., 1974. A new look at the statistical model identification. IEEE Transactions on Automatic Control 19, 716–723.

Amemiya, T., 1985. Advanced econometrics. Harvard University Press, Cambridge.

Arellano-Valle, R. B., Bolfarine, H., Lachos, V. H., 2005. Skew-normal linear mixed models. Journal of Data Science 3, 415–438.

Atienza, N., Garcia-Heras, J., Munoz-Pichardo, J. M., Villa, R., 2007. On the consistency of MLE in finite mixture models of exponential families. Journal of Statistical Planning and Inference 137, 496–505.

Benaglia, T., Chauveau, D., Hunter, D. R., Young, D. S., 2009. mixtools: An R

package for analyzing finite mixture models. Journal of Statistical Software 32, 1–29.

Boos, D. D., Stefanski, L. A., 2013. Essential statistical inference: theory and methods. Springer, New York.

Butler, S. M., Louis, T. A., 1992. Random effects models with non-parametric priors. Statistics in Medicine 11, 1981–2000.

Carnell, R., 2013. Triangle: provides the standard distribution functions for the triangle.
URL http://CRAN.R-project.org/package=triangle

Celeux, G., Martin, O., Lavergne, C., 2005. Mixture of linear mixed models for clustering gene expression profiles from repeated microarray experiments. Statistical Modelling 5, 243–267.

Chauveau, D., Hunter, D., 2013. ECM and MM algorithms for normal mixtures with constrained parameters. Tech. rep., HAL.

Chee, C.-S., Wang, Y., 2013. Estimation of finite mixtures with symmetric components. Statistics and Computing 23, 233–249.

Depraetere, N., Vandebroek, M., 2014. Order selection in finite mixtures of linear regressions: literature review and a simulation study. Statistical Papers 55, 871–911.

Eddelbuettel, D., 2013. Seamless R and C++ Integration with Rcpp. Springer, New York.

Efron, B., Tibshirani, R., 1993. An Introduction to the Bootstrap. Chapman and Hall, Dordrecht.

Galimberti, G., Soffritti, G., 2014. A multivariate linear regression analysis using finite mixtures of t distributions. Computational Statistics and Data Analysis 71, 138–150.

Gilbert, P., Varadhan, R., 2012. numDeriv: Accurate numerical derivatives. URL http://CRAN.R-project.org/package=numDeriv

Grun, B., Leisch, F., 2007. Fitting finite mixtures of generalized linear regressions in R. Computational Statistics and Data Analysis 51, 5247–5252.

Grun, B., Leisch, F., 2008. Flexmix version 2: finite mixtures with concomitant variables and varying and constant parameters. Journal of Statistical Software 28, 1–35.

Ho, H. J., Lin, T.-I., 2010. Robust linear mixed models using the skew t distribution with application to schitzophrenia data. Biometrical Journal 52, 449–469.

Keribin, C., 2000. Consistent estimation of the order of mixture models. Sankhya A 62, 49–65.

Kotz, S., Van Dorp, J. R., 2004. Beyond Beta: Other Continuous Families of Distributions with Bounded Support and Applications. World Scientific, Singapore.

Lachos, V. H., Ghosh, P., Arellano-Valle, R. B., 2010. Likelihood based inference for skew-normal independent linear mixed models. Statistica Sinica 20, 303–322.

Laird, N., 1978. Nonparametric maximum likelihood estimation of a mixing distribution. Journal of the American Statistical Association 73, 805–811.

Laird, N. M., Ware, J. H., 1982. Random-effects models for longitudinal data. Biometrics 38, 963–974.

Lange, K., 2013. Optimization. Springer, New York.

Lee, S. X., McLachlan, G. J., 2016. Finite mixtures of canonical fundamental skew t-distributions: the unification of the restricted and unrestricted skew t-mixture models. Statistics and Computing To appear.

Leroux, B. G., 1992. Consistent estimation of a mixing distribution. Annals of Statistics 20, 1350–1360.

McCulloch, C. E., Neuhaus, J. M., 2011. Misspecifying the shape of a random effects distribution: why getting it wrong may not matter. Statistical Science 26, 388–402.

McCulloch, C. E., Searle, S. R., 2001. Generalized, linear, and mixed models. Wiley, New York.

McLachlan, G. J., 1987. On bootstrapping the likelihood ratio test statistic for the number of components in a normal mixture. Journal of the Royal Statistical Society Series C 36, 318–324.

McLachlan, G. J., Krishnan, T., 2008. The EM algorithm and extensions, 2nd Edition. Wiley, New York.

McLachlan, G. J., Peel, D., 2000. Finite mixture models. Wiley, New York.

Meng, X.-L., Rubin, D. B., 1993. Maximum likelihood estimation via the ECM algorithm: a general framework. Biometrika 80, 267–278.

Miller, J. J., 1977. Asymptotic properties of maximum likelihood estimates in the mixed model of the analysis of variance. Annals of Statistics 5, 746–762.

Ng, S. K., McLachlan, G. J., Ben-Tovim, K. W. L., Ng, S. W., 2006. A mixture model with random-effects components for clustering correlated gene-expression profiles. Bioinformatics 22, 1745–1752.

Nguyen, H. D., McLachlan, G. J., 2016. Laplace mixture of linear experts. Computational Statistics and Data Analysis 93, 177–191.

Pinheiro, J. C., Bates, D., DebRoy, S., Sakar, D., R Core Team, 2014. nlme: Linear and nonlinear mixed effects models.
URL http://CRAN.R-project.org/package=nlme

Pinheiro, J. C., Bates, D. M., 2000. Mixed-effects models in S and S-PLUS. Springer, New York.

Pinheiro, J. C., Liu, C., Wu, Y. N., 2001. Efficient algorithms for robust estimation in linear mixed-effects models using the multivariate t distribution. Journal of Computational and Graphical Statistics 10, 249–276.

Potthoff, R. F., Roy, S. N., 1964. A generalized multivariate analysis of variance model useful espeically for growth curve problems. Biometrika 51, 313–326.

R Core Team, 2013. R: a language and environment for statistical computing. R Foundation for Statistical Computing.

Razaviyayn, M., Hong, M., Luo, Z.-Q., 2013. A unified convergence analysis of block successive minimization methods for nonsmooth optimization. SIAM Journal of Optimization 23, 1126–1153.

Redner, R. A., 1981. Note on the consistency of the maximum likelihood estimate for nonidentifiable distributions. Annals of Statistics 9, 225–228.

Schwarz, G., 1978. Estimating the dimensions of a model. Annals of Statistics 6, 461–464.

Song, P. X.-K., Zhang, P., Qu, A., 2007. Maximum likelihood inference in robust linear mixed-effects models using multivariate t distributions. Statistica Sinica 17, 929–943.

Stram, D. O., Lee, J. W., 1994. Variance components testing in the longitudinal mixed effects model. Biometrics 50, 1171–1177.

Titterington, D. M., Smith, A. F. M., Makov, U. E., 1985. Statistical analysis of finite mixture distributions. Wiley, New York.

Verbeke, G., Lesaffre, E., 1996. A linear mixed-effects model with heterogeneity in the random-effects population. Journal of the American Statistical Association 91, 217–221.

Verbeke, G., Molenberghs, G., 2000. Linear mixed models for longitudinal data. Springer, New York.

Weiss, L., 1971. Properties of maximum likelihood estimators in some nonstandard cases. Journal of the American Statistical Association 66, 345–350.